\documentclass[11pt]{article}

\newcommand{\hf}{\frac{1}{2}}

\newcommand{\xn}{x_{n}}

\newcommand{\e}{e^{i k_{0}Y}}                                      

\newcommand{\kim}{ k_{1}^{\mu}}                                      
\newcommand{\kom}{ k_{0}^{\mu}}                                      
\newcommand{\ki}{ k_{1}}

\newcommand{\kn}{ k_{n}}

\newcommand{\km}{ k_{m}}

\newcommand{\kt}{ k_{2}}                                             
\newcommand{\ko}{ k_{0}}                                             
\newcommand{\yim}{ Y_{1}^{\mu}}                                      
\newcommand{\yin}{ Y_{1}^{\nu}}                                      
\newcommand{\kin}{ k_{1}^{\nu}}  
\newcommand{\kir}{ k_{1}^{\rho}}                                    
\newcommand{\kon}{ k_{0}^{\nu}}
\newcommand{\kor}{ k_{0}^{\rho}}                                      
\newcommand{\ktm}{ k_{2}^{\mu}}                                      
\newcommand{\ytm}{ Y_{2}^{\mu}}                                      
                                                              
\newcommand{\lpp}{\mbox {$e^{i\int _{c} \alpha (t)                             
k(t) \partial _{z} X(z+t) dt +ik_{0}X}$}}

\newcommand{\gvk}{ e^{i\sum _{n }k_{n}Y_{n}}}

\newcommand{\p}{\partial}                                           
\newcommand{\pp}{\partial ^{2}}

\newcommand{\gvks}{ e^{i\sum _{n\ge 0 } k_{n}(t )\tY _{n}(t )}} 

\newcommand{\li}{ \lambda_{1}}                                    
\newcommand{\lt}{ \lambda_{2}}                                    
                                        
\newcommand{\al}{\alpha }                                             
 
\newcommand{\tY}{\tilde Y}                                 
\newcommand{\lan}{\langle}
\newcommand{\ran}{\rangle}

\newcommand{\la}{ \lambda }                                           
\newcommand{\be}{\begin{equation}}                                             
\newcommand{\br}{\begin{eqnarray}}                                             
\newcommand{\ee}{\end{equation}}                                               
\newcommand{\er}{\end{eqnarray}}

\renewcommand{\theequation}{\thesubsection.\arabic{equation}}

\begin{document}                                                               
\title{
\hfill\parbox{4cm}{\normalsize IMSC/2012/2/2\\
}\\
\vspace{2cm}
Loop Variables and Gauge Invariant Exact Renormalization Group
Equations for (Open) String Theory.
%\thanks{}%
}
\author{B. Sathiapalan\\ {\em                                                  
Institute of Mathematical Sciences}\\{\em Taramani                     
}\\{\em Chennai, India 600113}}                                     
\maketitle                                                                     

\begin{abstract}                                                               
An exact renormalization group equation is written down for the world sheet theory
describing the bosonic open string in general backgrounds.  Loop variable techniques
are used to make the equation gauge invariant. This is worked out explicitly up to level 3. 
The equation is quadratic in the fields and 
can be viewed as a proposal for a string field theory
equation.  As in the earlier loop variable approach, the theory  has one extra space  dimension and mass is obtained by dimensional reduction.  Being based on the sigma model RG, it is background independent.
 It is intriguing that in contrast to BRST string field theory,  the gauge transformations are not modified by the interactions up to the level calculated. The interactions  can be written in terms of  gauge invariant field strengths for the massive higher spin fields and the non zero mass is essential for this. This is reminiscent of Abelian Born-Infeld action (along with derivative corrections) for the massless vector field, which is also written in terms of the field strength.
\end{abstract}                                                                 
\newpage                                                                       
\section{Introduction} 

The renormalization group has been applied to the world sheet action for a string propagating in 
non-trivial backgrounds to obtain equations of motion [\cite{L}-\cite{T}]. One of the unsolved problems is
to write down gauge invariant (under space-time gauge transformations) RG equations for all the modes of the string
i.e. an exact renormalization group (ERG). This would then be equivalent to string field theory. The first systematic attempt to connect string field theory with the ERG was made in \cite{BM,HLP}.

 A generalization of this technique involving
Loop Variables has been used to give a partial solution to this problem
 \cite{BSLV,BSREV,BSCS}.  The free equation were written down. A version of the interacting equations were also written down.
There were a couple of noteworthy features :
The interacting equations were made to look exactly like the free equations (by employing the OPE and Taylor expansion
in the presence of a finite cutoff), and thus 
the mechanism of gauge invariance was very similar to that of the free theory. This is conceptually interesting. However it turns out that written in terms of
space time fields the gauge invariance of any equation involves contributions from all mass levels. This has the consequence
that one can either make the gauge invariance manifest, or the space time field structure manifest. 
Furthermore it is necessary to sum an infinite series of terms before the continuum limit (on the world sheet) can be taken. Thus it has not been possible thus far to write down a finite consistently truncated  set of equations in terms of space time fields where the gauge invariance is manifest.

In fact, in the approach to interactions advocated in \cite{BSREV}, it is possible to make field redefinitions, 
at any given order, that make the (space time theory) theory look free. However these field redefinitions
are singular when the cutoff is removed. This is  not surprising: As long as the cutoff is finite there are no poles in the string S-Matrix
and the theory is trivial.  The poles arise when the continuum limit is taken and then the theory is not free. So the continuum limit is crucial and it is thus important to write the theory in a way that allows this limit
to be taken. 

The  aim of this work is to write down a gauge invariant exact renormalization group (ERG). In the ERG
   the equations are quadratic.  The free part is the same as in the earlier works. The interaction
 terms are manifestly gauge invariant because they are written in terms of gauge invariant field strengths. In this aspect it differs from BRST string field theory.
 BRST string field theory \cite{SZ,WS,Wi2} also gives quadratic equations. However here (unlike in BRST string field theory) ,  the gauge transformations of the interacting theory is the same
as that of the free theory - at least up to level three, explicitly done in this paper. 
\footnote{It is also noteworthy that in the earlier approach of \cite{BSREV}, the gauge transformations
were affected by interactions.  This happens in the form of some trace constraints that are modified by the interactions. In the present
approach these constraints are not modified (at least up to level three).}

  Earlier work also addressed this problem from different angles. Some aspects of the finite
cutoff theory has been discussed  in \cite{BSPT, BSFC,BSLC} where it was shown that
if one keeps a finite cutoff, the proper time equation for the tachyon (which in this situation is related to the RG equation),  become quadratic. 
This is as expected both from
string field theory and also from the exact renormalization group [\cite{WK}-\cite{P}]. Some interesting aspects of the ERG have been discussed more recently in [\cite{BB1}-\cite{IIS}]. String field theory like equations derived from the ERG was written down in great detail in \cite{HLP} and it was also shown that the S-matrix was reproduced. In \cite{BSLC} 
it was also shown that one can make precise contact with light cone string field theory by keeping a finite cutoff.   In \cite{BSFC} 
it was also shown that if one wants to maintain gauge invariance
while maintaining a finite cutoff one needs to include all the massive modes in the proper time equation.
In this sense string field theory \cite{WS,SZ,Wi2,BZ} can be thought of as a way
of keeping a finite cutoff while maintaining gauge invariance. In \cite{BSERG,BSREV}  an exact ERG was written
down in position space and equations of motion derived. A proposal to make it gauge invariant was given. But
as mentioned above, it was not easy to write down equations in terms of space time fields, although in terms of loop variables  it was particularly easy. The present version is a modification that allows one to write down equations for space time fields with relative ease. 
One advantage of the RG approach is that the background about which one perturbs does not have to be conformal.
This is because the gauge invariance of the space time theory does not depend on any world sheet symmetry. This makes
the construction background independent. However in this paper only the equations of motion are given. There is no attempt
to construct an action. 
 
Another approach to off shell string theory
is the background independent approach of \cite{Wi} and further developed in \cite{LW,Sh,KMM}. 
The connection with the RG approach is discussed in \cite{Sh,KMM}. There is an elegant proposal for the action.  However there are 
problems in generalizing this proposal to general field configurations \cite{LW}. In \cite{BSERGTach}
a proposal was given for the action, which in fact reduced to the above results in the near on shell limit. However
the gauge invariant generalization using loop variables suffered from the same problems mentioned above (for the equations of motion).  
It would be interesting to see whether the techniques of this paper for gauge invariant interacting equations of motion can be used
to construct an action.

This paper is organized as follows: In Section 2 we review the derivation the ERG in position space and also the application to
 gauge fixed backgrounds - as one would in the "old covariant formulation" of string theory. This is a review of earlier work
  \cite{BSERG} and is included here for convenience. In section 3 we give a review of loop variables and derive the ERG using the loop variable
  language and 
we derive the gauge invariant version. This is substantially different from the approach in \cite{BSERG} as mentioned earlier in the introduction.
 Section 4 contains some explicit calculations for spin 2 and spin 3 fields. Section 5 contains some 
conclusions and  speculations.

\section{RG in Position Space}

In this section we derive the exact RG in position space. This is a repetition of Wilson's
original derivation \cite{WK}.   Note that usual discussions of the ERG use momentum space rather than position space.  
We start with  point particle quantum mechanics.  (This section is a review of ERG and also of some results from \cite{BSERG} 
which is reproduced here for convenience.)

\subsection{Quantum Mechanics}

Consider the Schrodinger equation
\be 
i\frac{\partial \psi}{\p t} =-\frac{\pp \psi }{\p y^2}
\ee  
for which the Green's function is $\frac{1}{\sqrt{2\pi (t_2-t_1)}}e^{i\frac{(y_2-y_1)^2}{2(t_2-t_1)}}$,
and change variables :$y = xe^\tau , it = e^{2\tau}$ and $ \psi ' = e^\tau \psi$ to get the
differential equation
\be
\frac{\p \psi '}{\p \tau} = \frac{\p}{\p x }(\frac{\p}{\p x} + x ) \psi ' 
\ee
The Green's function is:
\be
G(x_2, \tau _2; x_1 , 0) = \frac{1}{\sqrt{2\pi (1- e^ {-2\tau _2})}}
e ^{- \frac{(x_2 - x_1 e^{-\tau _2)^2}} {2(1-e^{-2\tau _2 )}}}
\ee

Thus as $\tau _2 \rightarrow \infty $ it goes over to $\frac{1}{\sqrt{2\pi}}e^{-\hf x_2^2}$. As $\tau _2 \rightarrow 0$ it goes to $\delta (x_1-x_2 )$. 
\[ \psi (x_2, \tau _2 ) = \int dx_1 G (x_2, \tau _2; x_1,0 )\psi (x_1, 0)
\]

So $\psi (x_2, \tau _2 )$ goes from being unintegrated
$\psi (x_1)$ to  completely integrated $\frac{1}{\sqrt{2\pi}}e^{-\hf x_2^2} \int dx_1 \psi (x_1)$. 
Thus consider
\be
 \frac{\p}{\p\tau } \psi (x_2, \tau) =  \frac{\p}{\p x_2}(\frac{\p}{\p x_2} + x_2) \psi (x_2, \tau )
\ee
with initial condition $\psi (x, 0) $
Thus  we can define $Z(\tau ) = \int d x_2 \psi (x_2, \tau )$, where $\psi $ obeys the above equation, we see that $\frac{d}{d\tau} Z =0$. When $\tau =0$ $\psi$ is the unintegrated $\psi (x,0)$. At $\tau = \infty$ it
is proportional to the integrated object $\int dx \psi (x,0)$. $Z(\tau )$ remains the same. Thus 
$\tau$  measures the extant to which $Z$ is integrated.

We will now  repeat this after taking the initial wave function as $e^{\frac{i}{\hbar}S[x ]}$ where
$x$ denotes the space-time coordinates. Then for $\tau = \infty$ $\psi  \approx \int {\cal D }x e^{iS[x]}$ is the integrated partition function. At $\tau =0$ it is the unintegrated $e^{iS[x]}$. $Z (\tau )$ is always the fully integrated
partition function. Following \cite{P}, we shall also split the action into a kinetic term and interaction term.
Thus  we write $\psi = e ^{-\hf x^2  f(\tau ) +L(x)}$  in the quantum mechanical case discussed above.

By choosing $a,b,B$ suitably  ( $b=2af , B = \frac{\dot f}{bf} $) in 
\[ \frac{\p \psi}{\p \tau} = B \frac{\p}{\p x} ( a \frac{\p}{\p x} + bx) \psi (x,\tau )
\] we get 
\be
\frac{\p L}{\p \tau} = \frac{\dot f}{2 f^2}[\frac{\pp L}{\p x^2} + (\frac{\p L}{\p x})^2 ]
\ee
Note that if $f = G^{-1}$ ($G$ can be thought of as the  propagator) then $\frac{\dot f}{f^2} = -\dot G$

\subsection{Field Theory}
We now apply this to a Euclidean field theory.

\be
\psi = e^{-\hf \int dz \int dz' X(z) G^{-1} (z,z') X(z') + \int dz L[X(z),X'(z)]}
\ee
Here $X'(z)=\p _z X(z)$. In general there could be higher derivatives $X''(z),X'''(z)...$. The equations can easily be generalized
to include those cases.
We apply the operator
\be   \label{27}
\int dz \int dz' B(z,z') \frac{\delta}{\delta X(z')} [\frac{\delta}{\delta X(z)} + \int b(z,z'')X(z'') ] 
\ee
to $\psi$ and require that this should be equal to $\frac{\p \psi}{\p \tau}$, as before,. 

We observe that 
\be \label{8}
\frac{\delta}{\delta X(z)} \int du ~L[X(u),X'(u)] = \int du~ [ \frac{\p L}{\p X(u)}\delta(u-z) +\frac{\p L}{\p X'(u)} \p _u \delta (u-z)]
\ee
and also 
\be \label{9}
\frac{\delta}{\delta X(z')}  \int du ~[ \frac{\p L}{\p X(u)}\delta(u-z)]=\int du~ [\frac{\pp L}{\p X(u)^2} \delta (u-z)\delta(u-z') + \frac{\pp L}{\p X(u)\p X'(u)}[\p _u \delta(u-z')]\delta(u-z)]\ee
and
\[ \frac{\delta}{\delta X(z')}\int du~[\frac{\p L}{\p X'(u)} \p _u \delta (u-z)]=\int du~ [\frac{\pp L}{\p X(u)\p X'(u)}[\p _u \delta(u-z)]\delta(u-z') +\]
\be \label{10}
\frac{\pp L}{\p X'(u)\p X'(u)}[\p _u \delta(u-z)][\p _u\delta(u-z')]]\ee

Adding (\ref{9})-(\ref{10}) and integrating by parts we get for the linear term:
\[
\frac{\delta^2}{\delta X(z)\delta X(z')} \int du ~L[X(u),X'(u)] =\frac{\pp L[X(z),X'(z)]}{\p X(z)^2}\delta(z-z') - \]
\be
\p_z [\frac{\pp L[X(z),X'(z)]}{\p X(z) \p X'(z)}]\delta(z-z') + \p_z\p_{z'} [\frac{\pp L[X(z),X'(z)]}{\p X'(z)^2}\delta(z-z')]
\ee

There is also a quadratic term. Again using (\ref{8}) in the form
\[ \frac{\delta}{\delta X(z)} \int du ~L[X(u),X'(u)] =  [ \frac{\p L}{\p X(z)} - \p _z\frac{\p L}{\p X'(z)} ]\]
we get for the quadratic term
\be
 [ \frac{\p L[X(z),X'(z)]}{\p X(z)} - \p _z\frac{\p L[X(z),X'(z)]}{\p X'(z)} ] [ \frac{\p L[X(z'),X'(z')]}{\p X(z')} - \p _{z'}\frac{\p L[X(z'),X'(z')]}{\p X'(z')} ]
\ee
We get the following five terms (all multiplied by $\int dz~\int dz'~B(z,z')$:
\[ (b-G^{-1})(z,z')
\]
\[ \Bigg( 
\frac{\pp L[X(z),X'(z)]}{\p X(z)^2}\delta(z-z') - 
\p_z [\frac{\pp L[X(z),X'(z)]}{\p X(z) \p X'(z)}]\delta(z-z') +\]\[ \p_z\p_{z'} [\frac{\pp L[X(z),X'(z)]}{\p X'(z)^2}\delta(z-z')]\]+
\[
 [ \frac{\p L[X(z),X'(z)]}{\p X(z)} - \p _z\frac{\p L[X(z),X'(z)]}{\p X'(z)} ] [ \frac{\p L[X(z'),X'(z')]}{\p X(z')} - \p _{z'}\frac{\p L[X(z'),X'(z')]}{\p X'(z')} ]\Bigg)
\]
 
\[   +\frac{\p L}{\p X(z)} (- \int G^{-1} (z',z'') X(z'') dz'')
\]
\[  +\frac{\p L}{\p X(z')} ( \int (b-G^{-1}) (z,z'') X(z'') dz'')
\]
\be - [(b- G^{-1}) X](z) [G^{-1} X](z') 
\ee
The first term is independent of $X$ and is therefore an unimportant overall constant. If we choose
$b = 2G^{-1}$, the third and fourth terms add up to zero (since $B(z,z')$ is symmetric under interchange of $z,z'$). 

Thus the second  term becomes
\[
\int dz \int dz' B(z,z') \Bigg( \Big(
\frac{\pp L[X(z),X'(z)]}{\p X(z)^2}\delta(z-z') - 
\p_z [\frac{\pp L[X(z),X'(z)]}{\p X(z) \p X'(z)}]\delta(z-z') +\]\[ \p_z\p_{z'} [\frac{\pp L[X(z),X'(z)]}{\p X'(z)^2}\delta(z-z')]\Big)\]+
\be
 [ \frac{\p L[X(z),X'(z)]}{\p X(z)} - \p _z\frac{\p L[X(z),X'(z)]}{\p X'(z)} ] [ \frac{\p L[X(z'),X'(z')]}{\p X(z')} - \p _{z'}\frac{\p L[X(z'),X'(z')]}{\p X'(z')} ]
\Bigg)
\ee

and the last term becomes:
\be
-\int dz \int dz' B(z,z') dz'' dz''' G^{-1} (z,z'') X(z'') G^{-1} (z', z''') X(z''') \psi
\ee

We can set (\ref{27})  equal to \[\frac{\p \psi}{\p \tau} = -\hf \int dz \int dz' X(z) \frac{\p G^{-1}}{\p \tau} (z,z') X(z') \psi
+ \int dz \frac{\p L}{\p \tau} \psi .\]
This ensures that $Z= \int {\cal D} X \psi $ satisfies $\frac{\p Z}{\p \tau} =0$.
If we now set $B = -\hf \dot G^{-1} (z,z')$ the equation for $\psi$ reduces to:
\[
\int dz \frac{\p L}{\p \tau} = -\int dz \int dz' \hf \dot G (z,z') \Bigg( \Big(
\frac{\pp L[X(z),X'(z)]}{\p X(z)^2}\delta(z-z') - 
\p_z [\frac{\pp L[X(z),X'(z)]}{\p X(z) \p X'(z)}]\delta(z-z') +\]\[ \p_z\p_{z'} [\frac{\pp L[X(z),X'(z)]}{\p X'(z)^2}\delta(z-z')]\Big)\]
\be  \label{RG}
+
\Big( [ \frac{\p L[X(z),X'(z)]}{\p X(z)} - \p _z\frac{\p L[X(z),X'(z)]}{\p X'(z)} ] [ \frac{\p L[X(z'),X'(z')]}{\p X(z')} - \p _{z'}\frac{\p L[X(z'),X'(z')]}{\p X'(z')} ]\Big)\Bigg)
\ee
 We can now take $\tau \approx ln~a$ and then this becomes easy to interpret as an RG equation diagrammatically
 \cite{P}:  the first curved bracket  in the RHS which is linear in $L$ represents
contractions of fields at the same point - self contractions within an operator. These can be understood as a prefactor multiplying normal ordered vertex operators. The second curved bracket represents contractions between fields at two different points - between two different operators. In terms of space-time fields, first term gives the free equations of motion and the second gives the interactions.

\subsection{ERG in the Old Covariant Formalism}

  We assume that the action is
\[
S = \int _0 ^R dz L[X(z)] = \int _0^Rdz \int dk [{\phi (k) \over a}e^{ikX(z)} + A_\mu (k) \p _z X^\mu e^{ikX(z)}
\]
\be \label{S}
+
\hf a S_2(k)^\mu \pp _z X^\mu e^{ikX(z)} + a S^{\mu \nu}(k)\p X^\mu \p X^\nu e^{ikX(z)}+...]
\ee
$a$ is a short distance cutoff. 
In order to implement the ERG we also need a specific form for $\dot G(z,z')=\dot G(z-z')$.
We need $\dot G (u)$ to be short ranged, otherwise the dimensionless ratio $R\over a$  will enter
in the equations. 
It is also a good idea to have analyticity so that one can perform Taylor expansions, which are required when
we do OPE's.
We will use the cutoff Green's function:
\be G(u) = \int {d^2k\over (2\pi )^2} \frac{e^{iku}e^{ - a^2 k^2} }{k^2}
\ee
This has a cutoff at short distances of $O(a)$ and at long distances reduces to the usual propagator.
We now apply the ERG (\ref{RG}) to the action S (\ref{S}).

The LHS gives
\be  \label{LHS}
\int dz \int dk ~[ {\beta _{\phi (k)} - \phi (k) \over a} \e + \beta _{A^\mu(k)} \p _z X^\mu(z)\e +...]  
\ee
where $\beta _g \equiv \dot g$.
The first term of the RHS gives
\be   \label{RHSI}
\int dz \int dk \hf (-k^2) {\e \over a} \phi (k)
\ee

The second term gives
\be   \label{RHSII}
\int dk_1 \int dk_2 {\phi (k_1) \phi (k_2) \over a^2} ({-k_1.k_2 \over 2}) \int _{-R}^{+R} du \dot G(u) e^{ik_1.X(z)}e^{ik_2.X(z+u)}
\ee
One can do an OPE for the product of exponentials to get\footnote{The OPE in terms of normal ordered is discussed
in the Appendix. We will use that to get the contribution of higher levels to lower level field equations. Here we do
not normal order.}
\[
e^{i(k_1+k_2)X(z) + ik_1 [u\p _z X + {u^2\over 2} \pp X + ...]}
\]

This gives 
\[
e^{i(k_1+k_2)X(z)} \int _{-R}^{+R} du ~\dot G(u) + ik_1 \p X e^{i(k_1+k_2)X(z)} \int _{-R}^{+R} du~ u \dot G(u)
\]
\[+ i k_1 {\pp X \over 2}  e^{i(k_1+k_2)X(z)} \int _{-R}^{+R} du ~ u^2 \dot G(u) +
{i k _\mu ik_ \nu \over 2} \p X^\mu \p X^\nu e^{i(k_1+k_2)X(z)} \int _{-R}^{+R} du~ u^2 \dot G(u) 
\]
It is easy to see that the first term of the OPE contributes to the tachyon equation:
\be
\beta _{\phi (k)} -\phi (k) = \phi (k) ({-k^2 \over 2}) - 
\hf \int dk_1 \phi (k_1) \phi (k-k_1) {k_1.(k-k_1)\over 2a}\int _{-R}^{+R} du~\dot G(u) 
\ee

Similarly the second term of the OPE gives the photon equation:
\be
\beta _{A^\mu (k)} =\int dk_1 {\phi (k_1) \phi (k-k_1) \over a^2} ({-k_1(.k-k_1) \over 2}) ik_1 ^\mu \int _{-R}^{+R} du~ u \dot G(u)
\ee

This gives the the tachyon contribution to the tachyon and photon equations of motion (EOM). More precisely these are contributions to the beta functions. The proportionality factor relating the beta function and the EOM  is the Zamolodchikov metric. Note that there is a dependence on $R/a$ in the coefficients, but from the form of the cutoff
chosen it is easy to see that there are always factors of $e^{-{R^2\over a^2}}$ accompanying the cutoff dependent terms and one can safely take the limit $R \rightarrow \infty$ when this dependence disappears. The equations then become independent of $a$ - even though  {\em $a$ is finite}. Thus the finite $a$ equations have the same form as the continuum equations. This is what is envisaged in the "improved" actions \cite{KS} or "perfect" actions \cite{PH}. Furthermore if the fields are tuned to the fixed point values, such that the beta function vanishes, then we have scale invariance at finite cutoff.

One can also include the contribution due to the photon field:
\[
\frac{\delta}{\delta X^\mu (z)} \int dz''~\int dk~ A_\nu (k)\p _{z''}X^\nu (z'') e^{ikX(z'')}
\]
\[
= \int dz'' \int dk A_\nu (k) [ \delta ^{\mu \nu} \p _{z''} \delta (z-z'') e^{ikX(z'')} + \p _{z''} X^\nu (z'')ik^\mu \delta (z-z'') \e ]
\]
\[ =\int dz'' \int dk A_\nu (k) [ -\delta ^{\mu \nu}  \delta (z-z'')ik^\rho \p _{z''}X^\rho e^{ikX(z'')} + \p _{z''} X^\nu (z'')ik^\mu \delta (z-z'') \e
\]
\[
=\int dz'' \int dk [-A^\mu (k)  ik^\nu  + ik^\mu A^\nu ] \p _{z''} X^\nu (z'')\delta (z-z'') \e
\]
\[
\frac{\delta ^2}{\delta X^\mu (z')X^\mu (z)}\int dz~L = \int dz'' \int dk \delta (z-z'') [ \p _{z''} \delta (z''-z') 
\underbrace{\delta ^{\nu \mu} [ -i k^\nu A^\mu + i A^\nu k^\mu ]}_{=0} \e \]
\be+ \p _{z''} X^\nu [ -ik^\nu A^\mu + ik^\mu A^\nu] ik^\mu \delta (z''-z')
\ee

The second term $\frac{\p L}{\p X^\mu(z)}\frac{\p L}{\p X^\mu (z')} $ becomes
\[\int dz'' \int dk [-A^\mu (k)  ik^\nu  + ik^\mu A^\nu ] \p _{z''} X^\nu (z'')\delta (z-z'') e^{ikX(z'')}\]
\[
\int dz''' \int dk [-A^\mu (k)  ik^\nu  + ik^\mu A^\nu ] \p _{z'''} X^\nu (z''')\delta (z'-z''') e^{ikX(z''')}
\]
Thus putting everything together we get
\[
\int dz~\int dz'~\dot G(z-z') \{\int dk [-A^\mu (k)  ik^\nu  + ik^\mu A^\nu ] ik^\mu \p _{z'} X^\nu (z')\delta (z-z') \e
\]
\be \label{Max}
+ \int dk \int dk' [-i k^{[\rho}A^{\mu ]}  ] [ -i{k'}^{[\sigma  } A^{\mu ]} ]\p _z X^\rho (z) \e \p _{z'}X^\sigma (z') e^{ik' X(z')}\}
\ee
The first
term is the usual Maxwell equation of motion. 
If we  assume analyticity of $\dot G(z-z')$) we can perform an OPE in the second term and re-express
as a sum of vertex operators for the various modes, just as in the case of the tachyon, above.  

The gauge invariance follows
 because of the integral over $z,z'$ which allows integration by parts. For the same reason, we have seen in loop variable calculations that if we do not introduce additional coordinates for higher gauge invariances,   one need not expect full gauge invariance in the EOM for higher (massive) modes.  The gauge invariance due to $L_{-1}$ is present  as the freedom to add total divergences in $z$. For the higher
gauge invariances due to $L_{-2}, L_{-3}...$ we need to be able to add total derivatives in some additional variables. This will be reviewed in the next section.
\section{Loop Variables and the ERG}

\subsection{Loop Variables}

 Loop variables
are useful when one wants to have a completely general background i.e. when all the massive modes are turned on as background. This is the domain of string field theory and  thus one is working with the full string field $\Phi [X(s)]$.
Thus for instance,
\be    \label{phi}
\Phi [X(s)] = \int ~[dk(s)]~ e^{i\int _c ~ds~k(s)X(s)} \Phi [k(s)] 
\ee

 In the limit $a\to 0$ we have a collection of vertex operators, 
all at the same point $z$.   
We have the following Taylor expansion:

\[
X(z+as) ~=~ X(z)~+~as \p _z X(z) ~+~ {1\over 2!} a^2s^2~\pp _z X(z) ~+~....
\]

We also assume that $k(s)$ can be expanded in a power series in $1/s$.
Thus 
\[
k(s) ~=~ k_0 ~+~ {\ki \over s} ~+~ {\kt \over s^2}~+~ ....
\]

Instead of (\ref{phi}), for later convenience we use the following definition of the loop variable:
\be     \label{L}
e^{i \ko  X(z) + 
i  \int _c ~ds ~k(s) \p _z X(z+as)}
\ee

When we expand the exponential in a power series we get the following terms:
\be  
e^{i \ko  X(z) + 
i  \int _c ~ds ~k(s) \p _z X(z+as)}= e^{i \ko  X(z)}[1+ i \kim \p _z X^\mu - \hf \kim \kin \p _z X^\mu \p _z X^\nu + \ktm \pp X^\mu +...]
\ee  

These are precisely the terms one writes down when one considers an open string in a general background, except that they are written in terms of loop variable momenta rather than space time fields. The connection becomes precise when 
we define space time fields in terms of loop variables.

The space time fields are defined by relations of the form:
\[
\lan 1 \ran = \phi
\]
\[
\lan \kim \ran = A^\mu
\]
\[
\lan \ktm \ran = S_2^\mu
\]
\be \label{defn}
\lan \kim \kin \ran = S_{11}^{\mu \nu}
\ee
where the $\lan ..\ran$ indicates an integration over some string functional of the $k_n$'s\footnote{The brackets $\lan \ran$ are also being used to denote the usual field theoretic correlations. It should be clear
from the context which is intended. }
Thus for instance,
\[
\lan \kim \ran \equiv \int \underbrace{{\cal D}k(s)}_{[d\ki d\kt ...d\kn ...]} \kim \Psi [\ko,  \ki, \kt,...\kn ..; \phi, A^\mu, S^{\mu \nu}...] = A^\mu (\ko)
\]
We have not done the integral over $\ko$, the usual space time momentum. Thus our fields are in momentum space.
This is only for convenience. One could integrate over $\ko$ and include $e^{i\ko .X(z)}$ in $\Psi$ if one wanted.

Note that the loop variable can also be written as

\be
\gvks
\ee

where \[ \tY _n \equiv {1\over (n-1)!}\p _z ^n X(z),~n>0;~~ \tY _0 \equiv X(z)
\]

\subsection{Gauge Invariance}

 We have seen that gauge invariance follows from the freedom to add derivatives. Thus if one
 is to retain all derivative terms one needs a {\em local} RG where the cutoff $a$ depends on $z$.
 This is illustrated below.
 
One can impose scale invariance by requiring that cutoff dependence vanish in expectation values  $\lan O_i \ran$.
This is equivalent to evaluating the effects of normal ordering.

Thus for the tachyon, ${1\over a}e^{ik.X} = e^{({k^2\over 2\pi }-1) ln ~a }:e^{ik.X}:$. 
Here :..: denotes normal ordering, so that $ \lan : O:\ran \equiv 0$  for all operators, except for the exponential : $\lan :\e: \ran \equiv1$. 
 ${d\over d ln~a } =0$ gives the equation of motion of the tachyon. Let us do this for the vector. But first we replace
 $a$ by $a e^{\sigma (z)}$ - this makes it local. $\sigma (z)$ can be thoought of as the Liouville mode.
\[
A^\mu (k ) \p _z X^\mu e^{ik .X} = - ik . A(k ) {\p _z \sigma \over 2\pi} :e^{ik .X}: e^{{k^2\over 2\pi} \sigma} + 
A^\mu (k) :\p _z X^\mu e^{ik.X}: e^{{k^2\over 2\pi} \sigma}
\]

Now we vary w.r.t $\sigma$, and integrate by parts, and then set $\sigma=0$ to find,
\[
(-k.A(k) k^\mu + k^2 A^\mu (k)) :\p_z X^\mu e^{ik.X}:
\] 

This is Maxwell's equation.  The $\p _z\sigma$
piece is crucial. The two physical state conditions $L_0=0$ and $L_1=0$ are combined into one equation. The gauge invariance of the equation ensures that this one equation is equivalent to both constraints.
This shows the role of
{\em local} scale invariance.

Suppose attempt to do the same thing for the massive modes of the form
$S_2 \p _z^2X e^{ik.X}$. 
One expects terms, for instance, of the form $k.S(k)\p_z^2 \sigma :e^{ik.X}:$. 
On varying w.r.t $\sigma$, the $z$ derivative acts twice on
  $e^{ik.X}$ and,
 we get terms of the form
$k.S(k) k^\mu k^\nu :\p _z X^\mu \p _z X^\nu e^{ik.X}:$. The resulting equation is cubic in derivatives and is not acceptable as an equation of motion. It doesn't get better at higher levels
\cite{BSLV}.    

So following \cite{BSLV} we introduce
additional variables $\xn , n>0$ that have the property that ${\p \over \p \xn }\approx \p _z ^n$.
On integrating by parts (and acting on $\e$), instead of getting $n$ powers of momenta, we get one. 

What could be the origin of these extra variables? They can be thought of as parametrizing diffeomorphisms of the loop variable. As it stands the loop variable is not diffeomorphism invariant but we can make it so by introducing an "einbein" along the loop. The modes of this einbein then provide the 
extra variables $\xn$. 

Let us consider the following loop variable:
\be   \label{84}
\lpp
\ee

$\al (s)$ is an einbein. Let us assume the following Laurent expansion:
\be
\al (s) ~=~ 1 ~+~ {\al _1 \over s}~+~{\al _2 \over s^2} ~+~ {\al _3 \over s^3}+...
\ee

Let us define 
\br
Y &~=~& X ~+~ \al _1 \p _z X ~+~ \al _2 \p_z^2 X ~+~
 \al _3 {\p _z ^3 X \over 2}~+~ ...~+~{\al _n \p _z^n X\over (n-1)!}~+~...\nonumber \\
&~=~& X ~+~ \sum _{n>0} \al _n \tY _n \\
Y_1 &~=~& \p _z X ~+~ \al _1 \p_z^2 X ~+~ \al _2 {\p _z ^3 X \over 2}~+~ ...~+~{\al _{n-1} \p _z^n X\over (n-1)!}~+~...\nonumber \\
...& & ...\nonumber \\
Y_m &~=~& {\p _z^m X\over (m-1)!} ~+~ \sum _{n > m}{\al _{n-m} \p _z^n X\over (n-1)!}\\
\er

Define $\al _0 =1$ so that the $>$ signs in the summations above can be replaced by $\ge$.

Thus
\be
\lpp = \gvk
\ee

where $Y_0=Y$.

 Introduce $\xn$ by the following:
\be
\al (s) = \sum _{n\ge 0} \al _n s^{-n} = e^{\sum _{m\ge 0} s^{-m} x_m}
\ee

Thus 
\br
\al _1 &=& x_1  \nonumber \\
\al _2 &=& {x_1^2 \over 2} + x_2 \nonumber \\
\al _3 &=& {x_1^3 \over 3!} + x_1x_2 + x_3
\er

They satisfy,
\be
{\p \al _n \over \p x_m} = \al _{n-m} , ~~ n\ge m
\ee

and thus
\be
Y_n = {\p Y\over \p x_n}
\ee

Let us  define  $\Sigma = \lan Y(z) Y(z) \ran$. This is a generalization of $\sigma$ to include the $\xn$ dependence,
 just as $Y$ is a generalization of $X$.
It is equal to the previous $\sigma$ when $\al (s) =1$.

Thus  the coincident two point functions become:
\br    \label{Sig}
\lan Y ~Y\ran &~=&~ \Sigma \nonumber \\
\lan Y_n ~Y\ran &~=&~ {1\over 2}{\p \Sigma \over \p x_n}   \nonumber \\
\lan Y_n ~Y_m \ran &~=&~  {1\over 2}({\pp \Sigma \over \p x_n \p x_m} - {\p \Sigma \over \p x_{n+m}})
\er

We normal order vertex operators as before to get:
\br   \label{LV}
\lpp &=& \gvk  \nonumber \\
&=& exp \{\ko ^2 \Sigma + \sum _{n >0} \kn .\ko  {\p \Sigma \over \p x_n} +  \nonumber \\
& & \sum _{n,m >0}\kn .\km {1\over 2}({\pp \Sigma \over \p x_n \p x_m} - {\p \Sigma \over \p x_{n+m}})\} \nonumber \\
& & :\gvk :
\er

Let us set  ${\delta \over \delta \Sigma}$ to zero (and also set $\Sigma =0$). 
As an illustration:

\[
{\delta \over \delta \Sigma} [
\kn .\km {1\over 2}({\pp \Sigma \over \p x_n \p x_m} - {\p \Sigma \over \p x_{n+m}})
] :e^{i\ko .Y}:
= :({1\over 2}i\kom i\kon Y_n^\mu Y_m ^\nu + i\kom  Y_{n+m}^\mu ) e^{i\ko .Y}:
\]  

If we now  collect all the coefficients of a particular vertex operator,
say $:Y_n ^\mu e^{i\ko .Y}:$, we get the free equation of motion. We can easily see that
they never contain more than two space-time derivatives. 

We can also understand gauge invariance as follows.  Having introduced an einbein we have to integrate
over all possible einbein fields, with a suitable measure ${\cal D} \al (s)$. It is this integration which allows
us to integrate by parts on the $\xn$.

Consider the following transformation: 
\be   \label{GT}
k(s) \to \la (s) k(s)
\ee

Clearly this is equivalent to $\al (s) \to \la (s) \al (s)$. But this is  just a change
of an integration variable. Assuming the measure is invariant this does nothing
to the integral. We choose ${\cal D} \alpha (s)$ to be $\prod _n d\xn $ and
 set $\la (s) = e^{\sum _m y_m s^{-m}}$. Then the gauge transformation (\ref{GT})
 is just a translation, 
$\xn \to \xn + y_n$ and leaves the measure invariant. Thus we conclude that
(\ref{GT}) is a gauge transformation. \footnote{$[L_{-n},Y_m]=m Y_{m+n}= m {\p\over \p \xn}Y_m$. This
gives the connection between the symmetry (Diff ($S^1$)) transformation in string theory and these gauge transformations.} 

We expand $\la (s)$ in inverse powers of $s$
\[
\la (s) = \sum _n \la _n s^{-n}
\]
and write (\ref{GT}) as 
\be    \label{GT1}
\kn \to \sum _{m=0}^{n} \la _m k_{n-m}
\ee

We set $\la _0 =1$.

We can interpret these equations in terms of space-time fields 
if we use (\ref{defn}), suitably  extended to include $\la$. Thus we
must assume that the string wave-functional is also a functional of $\la (s)$. 
Thus set
\br   \label{La}
\lan \la _1 \ran &~=&~ \Lambda _1 (\ko )\nonumber \\
\lan \la _1 \kim \ran &~=&~ \Lambda _{11}^\mu (\ko ) \nonumber \\
\lan \la _2 \ran &~=&~ \Lambda _2 (\ko )
\er

The gauge transformations (\ref{GT1}) in terms of space time fields are given by evaluating $\lan .. \ran$:

\br
A^\mu (\ko )~&\to &~ A^\mu (ko ) + \kom \Lambda _1 (\ko ) \nonumber \\
S^\mu _2 (\ko )~&\to &~ S^\mu _2 (ko ) + \kom \Lambda _2 (\ko ) + \Lambda ^\mu _{11} \nonumber \\
S_{11}^{\mu \nu} ~&\to &~ S^{\mu \nu}_{11} + k^{(\mu}_0 \Lambda _{11}^{\nu )}
\er  
These are the canonical gauge transformations for a  spin two field.
\footnote{Later a dimensional reduction will be done that make the fields massive.}

Now the gauge transformation parameters of higher spin fields
obey a certain tracelessness condition \cite{F,SH}.  We will see this below also.

When one actually performs the
gauge transformation it changes the
normal ordered loop variable by a total derivative in $\xn$ 
which doesn't affect the equation of motion.
Thus the gauge variation of the loop variable is a term of the form
${d\over d\xn } [A(\Sigma ) B]$, where $B$ doesn't depend
on $\Sigma$. The coefficient of $\delta \Sigma$ is obtained as
\[
\int ~~\delta ({d\over d\xn } [A(\Sigma ) B]) =
\int ~~ ({d\over d\xn}( {\delta A\over \delta \Sigma } \delta \Sigma ) B +
  {\delta A\over \delta \Sigma } \delta \Sigma {dB\over d\xn })
\]
\[
=\int ~~[ - {\delta A\over \delta \Sigma } {dB\over d\xn } +
 {\delta A\over \delta \Sigma } {dB\over d\xn }]\delta \Sigma =0
\]
Note that we have  integrated by parts.

Actually one finds on explicit calculation that the variation is  a total
derivative only if we use some identities that constrain the form of $\Sigma$. However we 
 would like to leave $\Sigma$ unconstrained
when we vary. Thus constraints have to be imposed elsewhere. It turns out that the terms that have to be put to zero are all of the form
\be
\la _n \km . k_p ...
\ee   

where ... refers to any other factors of $k_m$ \cite{BSLV}. 
Thus all traces
of gauge parameters have to be set to zero.

In \cite{BSLV} spin-2 and spin-3
are explicitly worked out.

 The gauge transformation (\ref{GT}) 
is a {\em scale transformation in space-time}. It is local
along the loop. This is suggestive of a {\em space time renormalization group} interpretation of the symmetry group of string theory as speculated in \cite{BSLV}. This speculation was the motivation for this approach. 
 
\subsection{Dimensional Reduction}

The equations that one obtains following the above steps give massless equations of motion.

In Section 2  the mass, being the dimension of the operator,
was obtained from the canonical dimension of operators. This is just the number of derivatives. In the RG
we introduce powers of $a e^{i\sigma}$ to make the derivatives dimensionless and so when we count
powers of $a$ we get the canonical dimension. In the new scheme we need to introduce it in a way
consistent with the gauge invariance of the massless theory. We simply do a Kaluza-Klein reduction
and  thus we must let the momentum $\kom$ be a 27-dimensional
vector rather than a 26-dimensional one. We will let $k_0^{26} \equiv q_0$
stand for the mass as in Kaluza-Klein theories but  we will assume that $q_0 ^2$ is a multiple
of $1\over R^2$ rather than letting $\ko$ be
multiples of $1\over R$.  The extra dimension brings an infinite set of auxiliary fields with it. This fortunately is just
what we need in string theory as shown in \cite{SZ}. There it was shown that one can get all the necessary auxiliary fields from the bosonized ghost - except that the the first oscillator mode was set to zero.

 We thus set $q_0$ to $\sqrt {(P-1)}$, where
$P$ is the engineering dimension of the vertex operator. Thus
for the tachyon $P=0$, for the vector $P=1$ etc. But  in our case
the first mode $q_1$ will not be set to zero identically because
that would violate gauge invariance. We will impose relations consistent with gauge invariance that allow us to get rid of $q_1$. These are
given below along with definitions of space-time fields in terms of loop variables:

{\bf Level 2}:
\[
\lan q_1 \ran =0 .
\]
\[
\lan q_1 q_1 \ran = \lan q_2 q_0 \ran = S_2 q_0~; ~\lan \la _1 q_1 \ran = \lan \la _2 q_0 \ran = \Lambda _2 q_0.
\]
\be	\label{DR2}
\lan q_1 \kim \ran = \lan \ktm q_0 \ran = S_2^\mu q_0.
\ee

This implies 
\be
 q_2 \rightarrow q_2+2 \lt q_0
\ee
{\bf Level 3}:

\[
\lan q_1\kim \kin \ran  = {1\over 2} \lan k_2^{(\mu} k_1^{\nu)} q_0\ran = \hf S_{21}^{(\mu \nu)}q_0\]
\[\lan q_1 q_1 \kim \ran  = \lan k_3^\mu q_0^2\ran = S_3^\mu q_0^2\]
\[
\lan q_1 \ktm  \ran = \lan 2 k_3^\mu q_0- q_2 \kim \ran = 2 S_3^\mu q_0 - S_{12}^\mu 
\]
\[\lan q_1 q_2\ran =  \lan q_3 q_0\ran =  S_3 q_0\]
\[ \lan q_1^3\ran  =\lan q_3 q_0^2\ran \]
\[
\lan \li q_1 \kim \ran  = \lan \hf  \lt \kim q_0 + \hf \li \ktm q_0\ran = (\hf \Lambda_{ 12}^\mu + \hf \Lambda _{21}^\mu)q_0\]
\[
 \lan \lt  q_1 \ran=  \lan 2 \la _3 q_0 -\li q_2 \ran = 2 \Lambda _3 q_0 - \Lambda _{21} \]
 \be \label{DR3}
 \lan \li q_1 q_1\ran = \lan \la _3 q_0^2\ran = \Lambda _3 q_0^2
 \ee

The gauge transformations for fields involving $q_2,q_3$ are modified to:
\be   \label{GT3}
\delta (q_2 \kim)= ({3\over 2} \lt \kim + \hf \li \ktm )q_0 + \li q_2 \kom ~~,~~~\delta q_3 = 3 \la _3 q_0
\ee
Note that correspondence with spin theory requires that $q_0^2 =1$  for the level-2 field and $q_0^2 =2$ for the level-3 fields. 
Relations of this type  enable us to get rid of $q_1$
completely. The form of the relations is such as to maintain gauge invariance.

We summarize the results for the gauge transformations of the massive spin-2 and spin 3 fields field:

{\bf Level 2}
\[
\delta S_{11}^{\mu \nu} = \ko ^{(\mu } \Lambda_{11} ^{\nu )}
\]
\[
\delta S_{2}^\mu  = \Lambda _{11}^\mu + \ko ^{\mu } \Lambda _2
\]
\be
\delta S_2 = 2 \Lambda _2 q_0
\ee

These are in the ``standard'' form, where the extra auxiliary fields
$S_2$ and $S_2^\mu$ can be set to zero to recover the Pauli-Fierz
equations for massive spin-2 fields. Further details can be found in
\cite{BSLV} and references therein.

{\bf Level 3}

The corresponding relations for spin3 are as follows:
\[
\delta S_{111}^{\mu \nu \rho} = \ko ^{(\mu} \Lambda _{111}^{\nu \rho )}
\]
\[ \delta S_{21}^{\mu \nu} = \Lambda _{111}^{\mu \nu} + \hf \ko ^{(\mu} (\Lambda _{12}+ \Lambda _{21})^{\nu )}+ \hf \ko ^{[\mu}(\Lambda _{12}-\Lambda_{21})^{\nu]}
\]
If we separate the symmetric and antisymmetric parts, $S_{21}^{\mu \nu} = S^{\mu \nu} +A^{\mu \nu}$, and $\Lambda _S ^\mu= \hf  (\Lambda _{12}+ \Lambda _{21})^{\mu }$ and
$\Lambda _A ^\mu= \hf  (\Lambda _{12}- \Lambda _{21})^{\mu }$, then
\[
\delta S^{\mu \nu} = \Lambda _{111}^{\mu \nu} + \ko^{(\mu}\Lambda _S^{\nu)}~~~;~~~ \delta A^{\mu \nu} =  \ko^{[\mu}\Lambda _A^{\nu]}
\]
\[\delta S_3^\mu = \Lambda_{21}^\mu + \Lambda _{12}^\mu + \kom \Lambda _3= 2\Lambda_S^\mu + \kom \Lambda_3
\]
$S_3^\mu$ is naturally associated with the symmetric tensor $S^{\mu \nu}$.
\[
\delta S_{12}^\mu = {3\over 2} \Lambda_{12}^\mu q_0 + \hf \Lambda _{21}^\mu q_0+ \kom \Lambda _{21}q_0
\]
The combination $S_3^\mu q_0- S_{12}^\mu$ undergoes the transformation
\[
\delta(S_3^\mu q_0- S_{12}^\mu)= \Lambda_A^\mu q_0 + \kom (\Lambda_{21}^\mu - \Lambda _3^\mu)\]
and is thus naturally associated with the antisymmetric tensor $A^{\mu \nu}$.
Finally,
\[
\delta S_3 = 3 \Lambda _3 q_0
\]

\subsection{ERG and Loop Variables}

We have already seen that what we refer to as the loop variable, integrated over $z$ i.e.  
\[
 \int dz ~{\cal D} \al (t) \lpp 
\] 
is actually the interacting part of the action expressed in terms of the loop variable momenta $\kn$:

 \be
 = \int \underbrace {[dz dx_1 dx_2...dx_n...]}_{[dz]}~~\gvk = \int [dz]~~L[Y(z,\xn), \frac{\p Y}{\p x_1}, \frac{\p Y}{\p x_2},...,\frac{\p Y}{\p \xn}]
 \ee
Thus the variable $z$ now stands for $(z,x_1,x_2,...,x_n,...)$. Furthermore when we have two points, $z,z'$, they will denote
the sets of variables: 
\[(z_A, x_{1A},x_{2A},....,x_{nA},...),(z_B, x_{1B},x_{2B},....,x_{nA},...)\] 
The integrals $\int dz$ in the ERG will be replaced
by $\int ... \int dz dx_{1A}dx_{2A}..dx_{nA}...$. Thus we will be allowed to integrate by parts on the $\xn$'s
exactly as in the case of the free string described above. We will see that the linear terms in the
ERG equation reproduce the free string equation and the quadratic term describes the interactions. The interactions will turn out to be 
not fully gauge invariant. The full gauge invariance requires a further modification described later below.

\subsection{Free Equations}

The free equations are obtained from the terms that are linear in $L$ in (\ref{RG}).  We have to extend the meaning of $X'(z)$ to
$Y_n= \frac{\p Y}{\xn}$ for all $n$. Thus $\frac{\pp L}{\p X^2}$ becomes $\frac{\pp L}{\p Y^2}$ in our new notation. Noting
that in (\ref{RG}), $G(z,z')=\lan X(z) X(z')\ran$, we see that in the gauge invariant version, $\delta (z-z')$ in the first term is actually $\delta (z-z') \prod _n \delta (\xn -\xn')$ and so 
$G(z,z)=\lan Y(z) Y(z)\ran = \Sigma $ in the notation of
Section 3.

The loop variable is 
\[
e^{i(\kom Y^\mu + \kim \yim + \ktm \ytm + k_3^\mu Y_3^\mu+...)}
\]
{\bf Level 1:}

Let us express $L$ in loop variable notation:
\be
L= (i \kim \yim )\e
\ee
The coefficient of $\yim$ in the linear part of the ERG is
\be    \label{3.55}
\int dz \dot G(z,z)[(i\ko)^2 i \kim \yim \e  - \frac{\p}{\p x_1}( i\ko . i k_1 \e )]= [-(\ko)^2 i \kim \yim   + ( \ko . k_1 i\kom \yim)] \e 
\ee
This is clearly Maxwell's equation.

Note that in the ERG one can integrate by parts and let the derivatives act on $\dot G(z,z)$. In this case the LHS  of (\ref{3.55}) can be written as: 
\be
 \int dz [ \dot G(z,z)[(i\ko)^2 i \kim \yim \e]  +\frac{\p \dot G(z,z)}{\p x_1}[( i\ko . i k_1 \e )]]
 \ee
If we let $ G(z,z) = \Sigma $ the similarity with (\ref{LV}) is clear. 

{\bf Higher Levels:}

One can similarly look at the contribution of 
$k_n Y_n \e$ in $L$. It gives a contribution:
\[ \frac{\p \dot G(z,z)}{\p x_n}[( i\ko . i k_n \e )] \]
to the linear term of the ERG. This is recognizable as $-\frac{\p \Sigma}{\p \xn} \ko .k_n \e$.

Similarly consider $ ik_n. Y_n ik_m.Y_m \e$ in the Lagrangian. The new contribution translated
to the loop variable notation is after integration by parts:
\[
\int dz~\int dz'~\dot G(z,z') \p _z \p _z' [\frac{\pp L}{\p X'(z)\p X'(z')}\delta (z-z')] = -\int dz~\int dz'~\frac{\pp \dot G(z,z')}{\p x_n \p x'_m} k_n.k_m \delta(z-z')
\]
  
The delta function implies that
 \[\int dz~\int dz'~\frac{\pp \dot G(z,z')}{\p x_n \p x'_m}\delta (z-z')=\int dz~\lan Y_n(z) Y_m(z)\ran = \int dz~\hf( \frac{\pp}{\p x_n \p x_m} - \frac{\p}{\p x_{n+m}})\Sigma\]
 This should be
compared with (\ref{LV}). Thus, as should have been expected, the linear part of the ERG reproduces the free equations
of motion obtained in the usual loop variable approach reviewed in Section 3. The gauge invariance has already been argued
in Section 3 and also explicitly demonstrated in earlier papers. We now turn to the interacting equations.

\subsection{Quadratic Terms}
 
 Letting $X(z),X(z')$ be $Y(z_A),Y(z_B)$ and $X'(z)$ stand for $\frac{\p Y}{\p x_{nA}}$ and $X'(z')$ stand for $\frac{\p Y}{x_{mB}}$, for the various $\xn$'s
 it is easy to see that each of the four quadratic terms stand for (after integrating by parts as in the linear case), respectively, terms of the form:
 \[ \ko(A) .\ko (B) \dot G(z_A,z_B),~~~k_{nA}.\ko (B) \frac{\p \dot G(z_A,z_B)}{\p x_{nA}},\]\[~~~~k_{nB}.\ko (A) \frac{\p \dot G(z_A,z_B)}{\p x_{nB}},~~~
 k_{nA}.k_{mB} \frac{\pp \dot G(z_A,z_B)}{\p x_{nA}\p x_{mB}}\] 
 
 The argument for gauge invariance works exactly as in the free case and involves showing that gauge transformations result in total derivatives.
 We need only worry about the fields labeled by $A$ as their transformation is completely independent of fields at $B$.
 Thus for instance: under $k_n(A)\rightarrow \la _{nA} \ko (A)$ , the term $k_{nA}.\ko (B) \frac{\p \dot G(z_A,z_B)}{\p x_{nA}}\e $ goes over
 to $\la _n(A) \frac{\p}{\p x_{nA} } [\dot G(z_A,z_B)]\ko (A).\ko (B)\e $.  Similarly
 the term \[ k_n(A) Y_n(A) \e \ko (A).\ko (B) \dot G(z_A,z_B)\]
 goes over to $\la _n(A) \frac{\p}{\p x_{nA}} [e^{i\ko (A).Y(A)}]  \ko (A).\ko (B) \dot G(z_A,z_B)$. Thus the total
 change is of the form $\la _n(A) \frac{\p}{\p x_{nA}} [\e  \ko (A).\ko (B) \dot G(z_A,z_B)]$, a total derivative.  This guarantees that the equation obtained as the coefficient
 of $\dot G(z_A,z_B)$, which will involve integrating by parts on $x_{nA}$, will be gauge invariant under $k_n(A)\rightarrow \la _{n}(A) \ko (A)$. (The reader is encouraged to verify this!) 
 
 However there is a problem with the lower invariances of the form $k_n(A)\rightarrow \la _{pA} k_{n-p} (A)$, for the term
 $k_{nA}.\ko (B) \frac{\p \dot G(z_A,z_B)}{\p x_{nA}}\e $ becomes 
 
 $\la _p(A) \frac{\p}{\p x_{nA}}  [\dot G(z_A,z_B)]k_{n-p} (A).\ko (B)\e $. Although this is actually
 equal to $\la _n(A) \frac{\pp}{\p x_{pA}\p x_{n-pA}}  [\dot G(z_A,z_B)]k_{n-p} (A).\ko (B)\e $, the result of integration by parts is clearly
 not the same. The first form gives one derivative of $x_{nA}$ on the remaining terms, with a sign reversal, whereas the second version gives 
 two derivatives of $x_{n-p,A}$ and $x_{pA}$, {\em without} a sign reversal. It is the second version that we need for invariance under $\la _p(A)$, whereas
 the first one gives invariance under $\la _{n}(A)$. 
 
 The resolution of this is to split $k_n$ into pieces, each of which transforms only under some of the
 $\la _m$. This will be described in the next section where explicit calculations are performed for level two and level three.
 
 To summarize this section, we have given the ERG in terms of loop variables. The linear part gives the gauge invariant free equation.
 The quadratic part gives the interacting part of the equation. However there is an issue regarding gauge invariance for which we have to find
 a solution. This is given explicitly for level two and three in the next section.
 
 \section{Examples}
 
In this section we set the tachyon to zero, since there is no gauge invariance associated with it.
 We have already worked out the level 1 results, which gives the free Maxwell equation. We now turn to
 level 2.
 \subsection{Level 2:}
\subsubsection{Linear Terms:}

 The Lagrangian is:
 \be
 L= i \kim \yim \e+i\ktm \ytm \e -  \hf \kim \kin \yim \yin \e
 \ee
 
 The various terms are:
 
 {\bf I.}
 
 \be
 \int \int  dzdz'~\dot G(z,z')[\frac{\pp L}{\p X(z)^2} \delta (z-z')] =\int dz~\dot G(z,z)[ i\ko .i\ko i\ktm \ytm - \hf i\ko .i\ko \kim \kin \yim \yin]  \e
 \ee
 
 {\bf II.}
 
 \[
 \int \int dz dz'~\dot G(z,z')[- \p _z [\frac{\pp L}{\p X(z) \p X'(z)}]\delta(z-z')]=
 \]\[ \int dz ~\dot G(z,z) [-\p_{x_2} [ ik_2 .i \ko \e] +
 \p _{x_1} [ i\ko . k_1 \kin \yin \e]]
 \]
 \be= \int dz~ \dot G(z,z)[ -i k_2.i\ko i \kom \ytm + i\ko .k_1 [ \kim \ytm + \kin \yin i\ko .Y_1]\e
 \ee
 {\bf III.}
 
 \[
 \int \int dzdz'~\dot G(z,z') \p _z \p _{z'} [ \frac{\pp L}{\p X'(z)^2} \delta(z-z')]
 \]
 As an intermediate step we can write this as 
 \[\int \int dz dz'~ \p_z\p_{z'}[\dot G(z,z')] \delta(z-z')  \frac{\pp L}{\p X'(z)^2}
 = \int dz \lan Y_1(z) Y_1(z)\ran  \frac{\pp L}{\p X'(z)^2}\]
 \[= \int dz~ \hf (\frac{\pp}{\p _{x_1}^2}- \frac{\p}{\p _{x_2}})[\dot G(z,z)] \frac{\pp L}{\p X'(z)^2}=
 \int dz~\dot G(z,z) \hf (\frac{\pp}{\p _{x_1}^2}+ \frac{\p}{\p _{x_2}})\frac{\pp L}{\p X'(z)^2}
\]
\be=-\int dz~ \dot G(z,z) k_1.k_1 (i\ko . Y_2 + \hf( i \ko .Y_1)^2)\e
\ee

We can collect the coefficients of $\ytm$:
\be  
- (\ko)^2 i\ktm + \ko .k_2 i\kom + i \ko .k_1 \kim - k_1.k_1 i\kom
\ee
The coefficients of $\yim \yin$ are:
\be   \label{lin11}
\hf \ko^2 \kim \kin - \hf k_0.k_1 k_1^{(\mu}k_0^{\nu )} + \hf \kom \kon k_1.k_1
\ee
They are written as massless equations in one higher dimension.
These are gauge invariant under $\ktm \rightarrow \ktm+ \lt \kom + \li \kim ~,~~\kim \rightarrow  \kim +\li \kom$. 

After  dimensional reduction they become
\[
- \ko^2 i\ktm + \ko .k_2 i\kom + i \ko .k_1 \kim - k_1.k_1 i\kom~=~0
\]
\[
\hf (\ko^2+q_0^2) \kim \kin - \hf k_0.k_1 k_1^{(\mu}k_0^{\nu )} - \hf q_0^2 k_2^{(\mu}k_0^{\nu )}+ \hf \kom \kon k_1.k_1 + \hf \kom \kon q_2q_0~=~0
\]
\be   \label{lin2}
- \ko^2 iq_2 + 2 \ko .k_2 iq_0  - k_1.k_1 iq_0~ =~ 0
\ee

The gauge transformation law for $q_2$ is $\delta q_2 = 2 \lt q_0$.

As explained earlier the canonical dimension of the operator is obtained as $q_0^2$ in this formalism. Thus where, in the OC formalism,
in the LHS of the ERG (\ref{RG}),i.e.  ${\partial L\over \p \tau}$, we had both the contribution of the canonical scaling, and the $\beta$-function (see (\ref{LHS})), now we need only the beta function. The canonical dimension that gives the tree level mass shows up in the RHS of the ERG in the form of the "anomalous" term $q_0^2$.\footnote{This is why in the loop variable formalism, string theory looks like a massless theory in one higher dimension.}

\subsubsection{Quadratic Terms}

The contribution of quadratic terms to level 2 can come from various sources. It can come from level 1 as well as level 2. 
We need to calculate $\frac{\p L}{\p X(z)} - \p _z \frac{\p L}{\p X'(z)}$ and thence
\[ \int \int dz_A~dz_B~\dot G(z_A,z_B)[\frac{\p L[X(z_A), X'(z_A)]}{\p X(z_A)} - \p _{z_A} \frac{\p L[X(z_A), X'(z_A)]}{\p X'(z_A)}]
\]\be
[\frac{\p L[X(z_B), X'(z_B)]}{\p X(z_B)} - \p _{z_B} \frac{\p L[X(z_B), X'(z_B)]}{\p X'(z_B)}]
\ee
The mechanism of gauge invariance discussed in the previous section suggests that under the gauge transformation
$\ktm (A) \rightarrow \ktm (A)+ \lt (A) \kom (A) + \li (A) \kim (A) ~,~~\kim (A) \rightarrow \li (A) \kom (A)$, if this expression
is to be invariant then each of the two factors in the product should be invariant. But one can check that this is not so. 
Consider the level 2 term
\[
\p _{x_1} \frac{\p L[Y(z_A), Y_1(z_A)]}{\p Y_1(z_A)} = -(\kim \kin Y_2^\nu + \kim k_1.Y_1 i \ko .Y_1)\e
\]
One of the terms in the gauge variation is $\li \kim \ko .Y_1 i \ko .Y_1 \e$. It can easily be checked that there is no term that can cancel this.
What ensures gauge invariance is that the gauge variation of every term in the Lagrangian should be a derivative of some lower
level term. Thus the $\li$ variation of level two should give $\li \p_{x_1} \ki .Y_1= \li  \ki. \p_{x_1}^2 Y$. Although $\p_{x_1}^2 Y=Y_2$
there is a difference between the two. This distinction is important because $\frac{\pp }{\p x_1^2}$ 
is not identically equal to $\frac{\p}{\p x_2}$ - it is only so
when acting on $Y$.  In particular one gets different results when integrating by parts. The end result
 now is gauge invariant. Thus we need to find a combination of loop variables that gives only $\li$ and this should be the coefficient of
 $\p_{x_1}^2 Y$. This combination is
  \[K_{11}^\mu \equiv \ktm - Q_2 \kom \equiv \ktm - (q_2 -{q_1^2\over 2q_0}) \kom
  \] 
  with gauge transformation:
 \[
 \delta K_{11}^\mu = \li \kim ~~~;~~~ \delta Q_2 \kom = \lt \kom
 \]
 If we use $q_1q_1 = q_2 q_0$
 then $Q_2 = \hf q_2$. Thus the strategy is to write $K_{11}^\mu \p_{x_1}^2 Y^\mu + Q_2 \kom \ytm$ instead of $\ktm \ytm$.

We work out the consequences of  this explicitly now. 

First modify the form of the ERG to accommodate second derivatives in $L[X,X',X'']$ and we get 
$\frac{\p L}{\p X(z)} - \p _z \frac{\p L}{\p X'(z)} + \p_z^2\frac{\p L}{\p X''(z)}$. Let us evaluate this for the  Lagrangian \footnote{Note that this Lagrangian is identical to the earlier one used for the linear part. The rewriting only has the effect of generating
a different set of terms when one integrates by parts. Thus the total derivatives that are being added or dropped are different. This
is thus a physically equivalent Lagrangian.}:
\be
L=[ i(\ktm - \hf q_2 \kom ) \frac{\pp Y^\mu}{\p x_1^2} + \hf i q_2 \kom \frac{\p Y^\mu}{\p x_2} - \hf \kim \kin \yim \yin]\e
\ee

\[
\frac{\p L}{\p Y^\mu} = [i\kom  i(\ktm - \hf q_2 \kom ) \frac{\pp Y^\mu}{\p x_1^2} + i\kom \hf i q_2 \kom \frac{\p Y^\mu}{\p x_2}-i\kom \hf \kim \kin \yim \yin]\e
\]
\[
\p _{x_1} \frac{\p L}{\p Y_1^\mu} = -\kim k_1.Y_2 \e - \kim k_1.Y_1 i \ko .Y_1 \e
\]
\[
\p _{x_2} \frac{\p L}{\p Y_2^\mu}=\hf i q_2 \kom i \ko .Y_2 \e
\]
\[
\p _{x_1}^2 \frac{\p L}{\p (\p _{x_1}^2Y^\mu)}= i(\ktm - \hf q_2 \kom )(i \ko .Y_2 + (i\ko .Y_1)^2)\e
\]
Thus 
\[
\frac{\p L}{\p X(z)} - \p _z \frac{\p L}{\p X'(z)} + \p_z^2\frac{\p L}{\p X''(z)}=[i\kom  i(\ktm - \hf q_2 \kom ) \frac{\pp Y^\mu}{\p x_1^2} + i\kom \hf i q_2 \kom \frac{\p Y^\mu}{\p x_2}-i\kom \hf \kim \kin \yim \yin]\e 
\]
\[+\Big(\kim k_1.Y_2 \e + \kim k_1.Y_1 i \ko .Y_1 \e \Big) -\hf i q_2 \kom i \ko .Y_2 \e\]
\be   \label{erg2}
+i(\ktm - \hf q_2 \kom )(i \ko .Y_2 + (i\ko .Y_1)^2)\e
\ee

In (\ref{erg2}) one can replace $ \frac{\pp Y^\mu}{\p x_1^2}$ by $\frac{\p Y^\mu}{\p x_2}$. This expression is gauge invariant as can easily be checked
explicitly.
The coefficient of $Y_2^\nu$ is :
\be	\label{V2}
V_2^{\mu \nu}\equiv [-\kom k_2 ^\nu  + \kim \kin -\ktm k_0^\nu  + q_2 \kom k_0^\nu  ]
\ee
The coefficient of $\yim \yin$ is:
\be	\label{V11}
V_{11}^{\rho \mu \nu}\equiv [-\hf k_0^\rho \kim \kin + \hf k_1^\rho (\kim \kon+ \kin \kom) -  (k_2^\rho - \hf q_2 k_0^\rho ) \kom \kon]
\ee

The $V$'s defined above are gauge invariant and are the analogues of the field strength $F^{\mu \nu}\equiv \kom \kin  - \kon \kim$ for the photon.
If we  define
\[
L_1^\mu(z)\equiv F^{\mu \rho}Y_1^\rho (z)e^{i\ko.Y(z)}
\]
and 
\[
L_2^\mu(z) \equiv [V_2^{\mu \rho}Y_2^\rho(z) + V_{11}^{\mu \rho \sigma}Y_1^\rho(z) Y_1^\sigma(z)]e^{i\ko.Y(z)}
\]
Thus the quadratic terms in the ERG takes the following form
\[
\int dz_A\int dz_B \dot G (z_A,z_B)\Big[(L_1^\mu(z_A) +L_2^\mu (z_A))(L_1^\mu(z_B) + L_2^\mu(z_B))\Big]=
\]

\[
\int dz_A\int dz_B \dot G (z_A,z_B) [F _{\rho \nu}Y_1^\nu (z_A) + V_{2 \rho \nu}Y_2^\nu (z_A) +
V_{11\rho \mu \nu}Y_1^\mu(z_A) Y_1^\nu (z_A)]e^{i \ko (A) .Y (z_A)}
\]
\be \label{quad2}
[F ^{\rho }_{~\alpha}Y_1^\alpha (z_B) + V_{2 ~ \alpha}^\rho Y_2^\alpha (z_B) +
V_{11 ~\alpha \beta}^\rho Y_1^\alpha(z_B) Y_1^\beta (z_B)]e^{i \ko (B) .Y (z_B)}
\ee

Before combining with the linear term (\ref{lin2}, \ref{lin11}) one needs to perform the OPE's in (\ref{quad2}).

\subsubsection{Operator Product Expansion}

Using the results of Appendix B we can perform the OPE's.  The linear term in the ERG evaluates the contribution
from self contractions within a vertex operator. So we can assume that the vertex operators are normal ordered for
the purposes of the calculation of the quadratic term. The contractions indicated by the general formula
in Appendix B for quadratic terms involves contraction between fields at different points and is different
from the self contractions of normal ordering.  Thus in performing the OPE
one can assume that they are normal ordered.

Now we can use the results of Appendix B to write down the OPE between these terms. Thus for instance to use the formulae there, 
$\kin, p_1^\nu$ there will stand for $ F^{\mu \nu}$ and $\kir \kin, p_1^\rho p_1^\nu$ will stand for $ V_{11}^{\mu \rho \nu}$. After all the substitutions are made one gets interacting equations of motion for the combined level 1, level 2
system (and level 0 if we include the tachyon). 

The full result involves a large number of terms and is not very illuminating. We give a sampling of some of the terms below:

{\bf Contribution of level 2 (massive spin 2) and level 1 (photon) to Maxwell's equation}
\[
\int dz_A \dot G(z_A,z_A) i\p_\nu F^{\mu \nu} + \int dz_A\int dz_B~\dot G(z_A,z_B)~[i\hf(\p_\rho \p _\sigma F^{\nu \mu}) V_{11}^{\nu \rho \sigma} (G_{1,0})^2
+
\]
\[
 i\hf (z_B-z_A)(\p_\la V_{11\nu}^{ \rho \sigma})(\p^\mu \p_\rho \p_\sigma F^{\nu \la})(G_{1,0})^2\]
\[
-i G_{1,0}^2 G_{0,1}(\p_\la V_{11}^{\al \rho \sigma})(\p_\rho \p_\sigma V_{11\al}^{\la \mu})+{i\over 4}(z_B-z_A) G_{0,1}^2G_{1,0}^2 (\p_\al \p _\beta V_{11}^{\la \rho \sigma})(\p_\rho \p_\sigma \p^\mu V_{11\la}^{\al \beta})+\]
\[
-i G_{1,1} F^{\nu \sigma} V_{11}^{\nu \sigma \mu} + i (z_2-z_1) G_{1,1}G_{1,0} \p ^\mu \p ^\rho F^{\nu \sigma} V_{11} ^{\nu \rho \sigma} +\]\[
i{(z_2-z_1)\over 2} G_{1,1}^2 ( \p^\mu V_{11}^{\nu \rho \sigma}) V_{11\nu \rho \sigma} + 2 i G_{1,1}G_{1,0} V_{11}^{\nu \rho \sigma} \p_\sigma V_{11 \nu \rho}^{~~~~\mu}\]
\[ - i(z_2-z_1) G_{1,1} G_{1,0}G_{0,1} (\p^\mu \p^\la V_{11}^{\nu \rho \sigma}) \p_\sigma V_{11 \nu \rho \la} +...]=0\]
The argument of the Green function, $z_A-z_B$, has been suppressed.  The three dots represent contribution from other fields.

{\bf Some contributions of level 2 (massive spin 2)  and level 1 to level 2 equation}
\[
\int dz_A \dot G(z_A,z_A)[ \hf (\pp -1 )S_{11}^{\mu \nu} -\hf \p_\rho \p^{(\nu}S_{11}^{\mu)\rho} + \hf \p^{(\mu} S_2^{\mu )} + \hf \p^\mu \p^\nu S_{11\rho}^\rho  - \hf \p ^\mu \p^\nu S_2] +
\]
\[
\int dz_A \int dz_B~\dot G(z_A,z_B)[V_{11}^{\la \rho \mu}(\p_\rho F^{\la \nu})G_{10}- (z_B-z_A)V_{11}^{\la \rho \sigma} (\p_\rho \p_\sigma \p^\nu F_\la ^\mu){G_{10}^2\over 2}+\p_\rho V_{11}^{\la \mu \nu} F_\la ^\rho {G_{0,1}\over 2}
\]
\[ - (z_B-z_A)\p_\sigma V_{11}^{\la \rho \mu} \p^\nu \p_\rho F_\la^\sigma (G_{1,0}G_{0,1})
+{(z_B-z_A)^2\over 4}(\p_\delta V_{11}^{\la \rho \sigma})(\p_\rho \p_\sigma \p^\mu \p^\nu F_\la^\delta )G_{1,0}^2 G_{0,1} \]
\[
+G_{1,1}V_{11}^{\sigma \rho \mu} V_{11 \sigma \rho}^{~~~~~\nu} -{(z_2-z_1)^2\over 2} G_{1,1}^2 ( \p^\mu \p^\nu V_{11}^{\la \rho \sigma}) V_{11 \la \rho \sigma} 
\]
\[-G_{1,0}^2 ( V_{11}^{\la \rho \sigma})(\p_\rho \p_\sigma V_{11\la} ^{\mu \nu}) - G_{1,0}^2 (\p_\sigma V_{11}^{\la \rho \mu})(\p_\rho V_\la ^{\sigma \nu})
+{(z_B-z_A)\over 2} G_{1,0}^2G_{0,1} (\p_\al V_{11}^{\la \rho \sigma})(\p^\nu \p_\rho \p_\sigma V_\la ^{\al \mu})
\]
\[ +{(z_B-z_A)\over 2} G_{1,0}G_{0,1}^2 (\p^\nu \p_\al V_{11}^{\la \rho \sigma})( \p_\rho \p_\sigma V_\la ^{\al \mu} )
-{(z_B-z_A)^2\over 2}(\p_\al \p_\beta V_{11}^{\la \rho \sigma})(\p^\mu \p^\nu \p_\rho \p_\sigma V_{11\la}^{\al \beta}) {G_{1,0}^2G_{0,1}^2\over 4}+
\]
\[
+(z_2-z_1)G_{1,1}G_{1,0} (\p^\nu \p^\rho V_{11}^{\la \sigma  \mu})V_{11 \la \sigma \rho} -(z_2-z_1) G_{1,1}G_{1,0} (\p^\nu V^{\la \rho \sigma})(\p _\sigma V_{11\la \rho}^{~~~~\mu})
\]
\[
-{(z_2-z_1)^2\over 2} G_{1,1}G_{1,0}G_{0,1} (\p^\tau V^{\la \sigma \rho} )( \p^\mu \p^\nu \p_\sigma V_{11\la \rho \tau})+...]=0\]

where
\[
V_{11}^{\mu\rho\sigma}=i[-{\p^\mu S_{11}^{\rho\sigma}\over 2} +{\p^{(\rho} S_{11}^{\sigma)\mu}\over 2} - \p^\rho \p^\sigma S_2^\mu + \hf \p^\mu \p^\rho \p^\sigma S_2 ]\]
\[
V_2^{\mu \nu} = - \p^{(\mu} S_2^{\nu)} + S_{11}^{\mu \nu} - \p^\mu \p^\nu S_2 \]
are the gauge invariant field strengths.
In the above only the contribution from $V_{11}$ is given.

We reproduce the gauge transformations of the fields
\[ \delta S_{11}^{\mu \nu} = \p^{(\mu}\Lambda _{11}^{\nu)} ~~~;~~~\delta S_2^\mu = \Lambda _{11}^\mu + \p^\mu \Lambda _2 ~~~;~~~\delta S_2 = 2 q_0 \Lambda _2 \]
Note that $q_0$ has been set to 1.

\subsubsection{Dimensional Reduction}
We need to comment on the role of the D+1 the coordinate and dimensional reduction.
As explained in Sec 4.1.1, the role of $q_0$ is the give a mass to the fields in accordance with string theory spectrum.
This requires that $q_0^2$ be set equal to the canonical dimension of the operator. The value of $q_0$ is thus fixed
when the free equations are written down.
Note that $V_{11}^{\mu \nu 5} = - V_{11}^{5 \mu \nu}={i q_0\over 2} V_2^{\mu \nu}$and $V_2^{\mu 5}=0$, where
'$5$' is symbolic for the 27th extra dimension - called $\theta$ in this paper. 

 Thus we need
$G^{\theta \theta}(z,z) = G^{XX}(z,z)$, in order that the anomalous dimension come out as $k_{0\mu} \kom = \ko ^2 + q_0^2$ in the linear part of the ERG.  However we {\em do not want contributions from $q_0$ in correlation functions between
vertex operators at different locations.} This would affect the pole structure of the S-matrix. Thus we want $G^{\theta \theta}(z,z') \rightarrow 0$ when $z\neq z'$. This can be achieved by making the $\theta$ coordinate massive - with
a mass of the order of the UV cutoff. Thus we take
\be    \label{theta}
\lan \theta(z) \theta (z')\ran = \int d^2 q e^{iq .(z-z')} {1\over q^2 +m^2 } 
\ee
with $m \approx {1\over a}$ where $a$ is the short distance cutoff or the lattice spacing.

Note that this implies that there is no translation invariance in the $\theta$ direction and there is no $q_0$-momentum conservation. The value of $q_0$ when it occurs in a field is fixed once and for all by the linearized theory. In computations, this means that in the ERG, the linear term gets a contribution from $\theta$ contractions in the normal ordering, but in the quadratic
term the sum over $\mu$ does not include $\theta$. $\theta$ will continue to appear in the vertex operators for external states.

The  propagator (\ref{theta}) would violate conformal invariance on the world sheet. However it does not affect the 
S-matrix for physical states or the space-time gauge invariance of the theory. The S- matrix is not affected because
it has been argued \cite{BSOC} that the world sheet  interaction Lagrangian for {\em physical} external states in the loop variable formalism reduces to that of the Lagrangian of the "Old Covariant" formalism with physical state constraints.\footnote{We caution that the demonstration has been explicitly done only for the second and third massive levels.} Therefore if $\theta$ does not affect the correlation functions, the equivalence of the S-matrix follows. 
Space-time gauge invariance is not affected because this is built into the loop variable formalism and does not rely on world sheet symmetries.

The interactions are manifestly invariant under the {\em same} gauge transformations that leave the linear term invariant, i.e.
{\em the gauge transformation is not modified by the interactions.}  This is different from BRST string field theory where
the gauge transformations are modified by the interactions and only the full equations of motion are invariant. In this sense
we have an Abelian theory rather than a non-Abelian theory. It is possible that some field redefinitions in the BRST string field theory
formulation will make it equivalent to this one. We also note that if we introduce Chan-Paton factors, the gauge transformations
as well as the interactions will be modified in this formalism also.

\subsection{Level 3}

\subsubsection{Linear Terms}

The Lagrangian is:
\be
L= [ik_3^\mu Y_3^\mu - \ktm \kin \ytm \yin - {i\over 3!} \kim \kin \ki ^\rho \yim \yin Y_1^\rho]\e
\ee 
{\bf I.}

The first term in the ERG is $\int dz \dot G(z,z) {\pp L\over \p X(z)^2}$ which gives
\[
\int dz \dot G(z,z) (-\ko ^2)[ik_3^\mu Y_3^\mu - \ktm \kin \ytm \yin - {i\over 3!} \kim \kin \ki ^\rho \yim \yin Y_1^\rho]\e
\]
{\bf II.}

The next term is: $ -\int dz \dot G(z,z)\p_z {\pp L\over \p X(z)\p X'(z)}$.
Using
\[
 {\pp L\over \p Y_0^\mu(z)\p \yim(z)}= i\kom [-(\kt . Y_2) \kim - {i\over 2!} \kim (k_1 .Y_1)^2]\e=[-(\kt.Y_2)i\ko.\ki + \hf \ko .\ki (\ki.Y_1)^2]\e
 \]
 we get:
\[
\p_{x_1} {\pp L\over \p Y_0^\mu \p \yim (z)}=\]\[[-k_2.Y_3 i\ko.\ki + \ko.\ki k_1.Y_1 k_1 .Y_2 ] \e+ i\ko.Y_1 [-(k_2.Y_2)i\ko .\ki +\hf \ko.\ki (\ki.Y_1)^2]\e
\]
Similarly
\[
{\pp L\over \p Y_0^\mu(z)\p \ytm(z)}=i\kom[-(\ki .Y_1)\ktm ]\e = -i (\ko .\kt) (\ki .Y_1) \ko.Y_2\e
\]
\[
\p_{x_2}{\pp L\over \p Y_0^\mu(z)\p \ytm(z)}=[-i(\ko.\kt)\ki.Y_3 +(\ko.\ki) (\ki.Y_1)( \ko.Y_2)]\e
\]
and
\[
\p_{x_3}{\pp L\over \p Y_0^\mu(z)\p Y_3^\mu(z)}=-i\ko.k_3 (\ko.Y_3)\e
\]
{\bf III.}

The last term is $\int \int dz dz' ~\p_z \p_{z'} \dot G(z,z') \delta(z-z'){\pp L\over \p _z X'(z)^2}$ which gives
\[
\hf ({\pp \over \p_{x_1}^2}-{\p\over \p_{x_2}})\dot G(z,z){\pp L \over \p \yim \p \yim} +({\pp \over \p_{x_1}\p_{x_2}}-{\p\over \p_{x_3}})\dot G(z,z){\pp L \over \p \yim \p \ytm}
\]
\[
{\pp L \over \p \yim \p \yim}=-i \ki.\ki (\ki.Y_1) \e ~~~~;~~~~~{\pp L \over \p \yim \p \ytm}=-\kt.\ki \e
\]
Integrating by parts we get
\[
\hf ({\pp \over \p_{x_1}^2}+{\p\over \p_{x_2}})[-i \ki.\ki (\ki.Y_1) \e]=\]\[-{i\over 2}[\ki.\ki 2 ( \ki.Y_3  +  \ki .Y_2 i\ko.Y_1+ \ki.Y_1 \ko.Y_2)-
{i\over2} \ki.\ki \ki.Y_1 (i\ko. Y_1)^2]\e
\]
\[
({\pp \over \p_{x_1}\p_{x_2}}+{\p\over \p_{x_3}})[-\kt.\ki \e]=-\kt.\ki [2 i \ko.Y_3 + i\ko.Y_1i\ko.Y_2 ]\e
\]

Adding the contributions of {\bf I, II, III} we get
\[
{\bf Y_3^\mu}[ -\ko^2 ik_3^\mu + i\ko.\ki \ktm + i\ko.\kt \kim + i\ko.k_3 \kom - i\ki.\ki \kim - 2i\ki.\kt \kom ]
\]
\[
+ {\bf Y_2^\mu Y_1^\nu}[\ko^2 \ktm \kin -\ko.\ki \kim\kin -\ko.\ki \ktm \kon -\ko.\kt \kom \kin +\ki.\ki \kom \kin + \ki.\ki \kim \kon + \ki .\kt \kom \kon]
\]
\be
+ {\bf \yim \yin Y_1^\rho}[{i\over 3!}\ko^2 \kim \kin \kir - {i\over 2} \ko.\ki \kim \kin \kor + {i\over 2} \ki.\ki \kim \kon \kor ] =0
\ee
These are the linear Spin 3 equations. They are gauge invariant under
\[
k_3^\mu  \rightarrow k_3^\mu+ \la _3 \kom + \lt \kim + \li \ktm ~~,~~~~~ \ktm \rightarrow \ktm+ \lt \kom + \li \kim ~,~~\kim \rightarrow \kim +\li \kom
\]

 Note that they are written as massless higher dimensional equations. Mass can be introduced by dimensional reduction in the usual way. 
 When this is done and the substitutions given in (\ref{DR3}) are made we get \footnote{The symmetrization symbols imply adding the permutations required 
 for complete symmetry. Thus for eg. $\kim \kon \kor$ requires three terms whereas $\ktm \kin \kor$ requires six terms.}:
 \[
{\bf Y_3^\mu}[ -\ko^2 k_3^\mu + \ko.\ki \ktm + \ko.\kt \kim + \ko.k_3 \kom - \ki.\ki \kim -2\ki.\kt \kom -5 q_3 q_0 \kom] +
 \]
 \[
{\bf \ytm \yin}[ \hf \ko^2 k_2^{(\mu} k_1^{\nu)} + \hf(\ko^2 +q_0^2) k_2^{[\mu} \ki ^{\nu]}  -q_0^2 k_3^{[\mu} k_0^{\nu]} +\]\[
q_0q_2 k_1^{[\mu}k_0^{\nu]} -
 \ko.\ki (\kim \kin + \ktm \kon) - \ko.\kt \kom \kon + \ki.\ki \ko^{(\mu}\ko^{\nu)} +3 q_3 q_0 \kom \kon]+\]
 \be   \label{lin3}
{\bf  \yim\yin Y_1^\rho} [(\ko^2+q_0^2){\kim \kin \kir\over 3!}- \ko.\ki {\ki^{(\mu}\kin \ko^{\rho)} \over 3!}-  q_0^2 {k_2^{(\mu} \kin \ko^{\rho)}\over 12} +
 \ki.\ki {\ki^{(\mu} \kon \ko^{\rho)}\over 3!} + q_0^2 {k_3^{(\mu} \kon \ko^{\rho)}\over 3!} ]  
 =0
\ee
 
 The tracelessness condition becomes $\li \ki.\ki + \li q_1 q_1 = \li \ki.\ki + \la _3 q_0^2=0$ and is required for gauge invariance of the above equations. 
 The gauge transformations include the ones in (\ref{GT3}).
 
 \subsubsection{Quadratic Terms}

The first step is to find  combinations of variables such that their gauge variations are derivatives of the lower level terms. We write 
the Level 3 term $k_3^\mu Y_3^\mu$ as 
\[
K_3^\mu {\p Y^\mu\over \p x_3} + K_{21}^\mu {\pp Y^\mu \over \p x_2 \p x_1} + K_{111}^\mu {\p^3 Y^\mu\over \p x_1^3}
\] whose gauge variations are
\[
\delta K_3^\mu = \la_3 \kom ~~,~~\delta K_{21}^\mu = \lt \kim + \li Q_2 \kom ~~~,~~~\delta K_{111}^\mu = \li K_{11}^\mu
\]
This gives us:
\[
\delta( K_3^\mu {\p Y^\mu\over \p x_3}) = \la _3 {\p\over \p x_3} (\ko.Y)~~,~~~\delta(K_{21}^\mu {\pp Y^\mu \over \p x_2 \p x_1})=\lt {\p\over \p x_2}(\ki. Y_1) + \li {\p \over \p x_1} (Q_2 \ko. Y_2)
\]
\[
\delta (K_{111}^\mu {\p^3 Y^\mu\over \p x_1^3})= \li {\p \over \p x_1} (K_{11}. {\pp Y \over \p x_1^2})
\]

We give the solution below:
\[
q_0K_3^\mu = \hf[q_0 k_3^\mu -(q_1\ktm +q_2\kim - {q_1^2\kim\over q_0} +{q_1^3\kom\over 3 q_0^2}-q_3\kom)]
\]
\[
q_0 K_{21}^\mu = q_2\kim - {q_1^2\over 2q_0}\kim
\]
\[
q_0K_{111}^\mu= \hf(q_0k_3^\mu + q_1\ktm -q_2\kim) -({q_3\over 2} - {q_1^3\over 6 q_0^2})\kom
\]
Note that the sum of the three is $q_0k_3^\mu$ as it should be.

We use the above to write the Level 3 Lagrangian, $L$:
\[
L= \Big[ iK_3^\mu Y_3^\mu +  iK_{21}^\mu {\pp Y^\mu \over \p x_2 \p x_1} + iK_{111}^\mu {\p^3 Y^\mu\over \p x_1^3} 
-K_{11}^\mu \kin {\pp Y^\mu\over\p x_1^2}{\p Y^\nu\over \p x_1} 
\]
\be  \label{L3}
 -Q_2 \kom \kin {\p Y^\mu\over \p x_2} {\p Y^\nu\over \p x_1} - {i\kim \kin \kir\over 3!}
\yim \yin Y_1^\rho\Big]\e
\ee

The quadratic term calculated in Appendix A is written in terms of $L(z)$ defined below:
\[ L_3^\mu(z)\equiv \Big[V_3^{\mu \nu}Y_3^\nu(z) + V_{21}^{\mu \rho \sigma} Y_2^\rho(z) Y_1^\sigma(z) + V_{111}^{\mu \lambda \rho \sigma} Y_1^\lambda(z) Y_1^\rho (z)Y_1^\sigma (z)\Big] e^{i\ko.Y(z)}
\]
where
\[
V_3^{\mu \rho}=-\kom[K_3^\rho + K_{21}^\rho + K_{111}^\rho] + \kim [K_{11}^\rho + Q_2 \kor] + \kom Q_2 \kir - K_{11}^\mu \kir - K_{21}^\mu \kor + K_{111}^\mu \kor + K_3^\mu \kor
\]
\[
V_{21}^{\mu \rho\sigma} = i\Big[ -\kom K_{11}^\rho \kir +  \kim K_{11}^\rho \ko^\sigma + \kim Q_2 \kor \ko^\sigma + \kim \kir \ki^\sigma
-2 K_{11}^\mu \kir \ko^\sigma -K_{11}^\mu \kor \ki^\sigma - K_{21}^\mu\kor \ko^\sigma + 3 K_{111}^\mu \kor \ko^\sigma \Big]
\]
\[
V_{111}
^{\mu \la \rho \sigma}= 
{1\over 3!} 
\kom k_1 ^\la \kir \ki ^\sigma -
 {1\over 3!}\kim \ki ^{(\lambda} \kir \ko^{\sigma)} + 
{1\over 3} K_{11}^\mu \ki ^{(\lambda} \kor \ko^{\sigma)} - K_{111}^\mu \ko^{\lambda} \kor \ko^{\sigma}
\]
$L_3^\mu(z)$ is a gauge invariant field strength for the massive level 3 fields. Note that the non-zero mass ($q_0$) is crucial for being able to construct such an object.  

We can now eliminate $q_1$ as before using (\ref{DR3}).
After eliminating $q_1$ they become:
 \[
 q_0K_3^\mu  ={q_3\over 3}\kom
 \]
\[
q_0K_{111}^\mu= {3\over 2}k_3^\mu q_0 - q_2 \kim - {q_3\over 3}\kom
\]
\be
q_0K_{21}^\mu= \hf(2q_2\kim- q_0k_3^\mu)
\ee

The quadratic term can thus be written in a manifestly gauge invariant way as:
\be
\int \int dz_A~dz_B~\dot G(z_A,z_B) (L_1^\mu(z_A) + L_2^\mu(z_A) + L_3^\mu(z_A))(L_1^\mu(z_B) + L_2^\mu(z_B) + L_3^\mu(z_B))
\ee
where we have included all the fields from lower levels\footnote{Except for the tachyon.}.
 Once we have a gauge invariant equation we can set $x_n=0$
and these vertex operators reduce to standard ones. An OPE has to be then performed in the same way as was done for level 2 before we can combine this with the linear term.
As mentioned earlier, it is very interesting that the gauge transformation is the same linear transformation of the free theory.

As in the case of level 2 these equations can be converted to space time form after the OPE's are performed.
However since the result is not particularly illuminating we do not do it here. We hasten to add that the
method while tedious is quite straightforward as we have seen in the level 2 case.

\subsection {Extension to Level 4}

The extension to level 4 is outlined to illustrate the general pattern. In practice the algebra may be tedious and has
not been attempted. The pattern is easy to easy once we list the operators and gauge transformation at each level:

{\bf Level 0}

\[\ko.Y\]

{\bf Level 1}
\[\kim {\p Y\over \p x_1}~~~\rightarrow \la _1 {\p (\ko.Y)\over \p x_1}  \]

{\bf Level 2}
\[ K_{11}{\pp Y\over \p x_1^2}  ~~\rightarrow  \la _1  {\p\over \p x_1} ( \ki {\p Y\over \p x_1}) \]
\[+ Q_2 \ko {\p Y\over \p x_2} ~~\rightarrow \la _2 {\p \over \p x_2} (\ko Y) \]

{\bf Level 3}
\[ K_3 {\p Y\over \p x_3} ~~\rightarrow \la _3 {\p \over \p x_3} (\ko Y) \]
\[+ K_{21} {\pp Y\over \p x_1 \p x_2} ~~~\rightarrow \la _2 {\p \over \p x_2}( \ki {\p Y\over \p x_2}) + \la _1 {\p \over \p x_1}( Q_2\ko {\p Y\over \p x_2})\]
\[+ K_{111} {\p^3 Y\over \p x_1^3} ~~~\rightarrow \la _1 {\p \over \p x_1} (K_{11} {\pp Y \over \p x_1^2})\]

From the above pattern for level 4 we need to find:

{\bf Level 4}

\[ K_4 {\p Y\over \p x_4} ~~~\rightarrow \la _4 {\p \over \p x_4 }(\ko Y)\]
\[+ K_{31} {\pp Y\over \p x_3 \p x_1} ~~~\rightarrow \la _3 {\p \over \p x_3}(\ki {\p Y\over \p x_1}) + \la _1 {\p \over \p x_1}( K_3 {\p Y\over \p x_3}) \]
\[+ K_{22} {\pp Y\over \p x_2^2} ~~~\rightarrow \la _2 {\p \over \p x_2} ( Q_2\ko {\p Y\over \p x_2})\]
\[ + K_{211} {\p^3 Y\over \p x_2 \p x_1^2} ~~\rightarrow \la _2 {\p \over \p x_2} K_{11} {\pp Y\over \p x_1^2}+ \la _1 {\p \over \p x_1} (K_{21} {\pp Y\over \p x_2 \p x_1})\]
\[+ K_{1111} {\p ^4 Y\over \p x_1^4} ~~\rightarrow \la _1 {\p \over \p x_1}(K_{111}{\p^3 Y\over \p x_1^3})\]

Thus we need to find combinations of the $\kn $ and $q_n$ such that
\[ \delta K_4^\mu = \la _4 \kom~~~,~~~\delta K_{31}^\mu = \la _3 \kim + \la _1 K_3^\mu ~~~,~~~ \delta K_{22}^\mu = \la _2 Q_2 \kom \]\[K_{211}^\mu = \la _2 K_{11}^\mu + \la _1 K_{21}^\mu  ~~~,~~ K_{1111}^\mu ~~\rightarrow \la _1 K_{111}^\mu \]

\subsection{Equivalence with String Theory}

In deriving gauge invariant equations the main ingredient was the freedom to add total derivatives in $\xn$. One can ask
whether the theory is still equivalent to string theory. Is the S-matrix defined by this theory the same as that of string theory? It has been shown that for the free theory, one can map the fields, gauge transformations and constraints to those of the old covariant formalism - for level two and three \cite{BSOC}. Since the
interactions are generated by calculating correlation function of vertex operators and this procedure is mathematically the same in both cases, the interacting theory should give the same physical results. A formal proof of this equivalence however has not been addressed.

\section{Summary and Conclusions}

In this paper we have written down the exact renormalization group (ERG) for the world sheet action describing
an open string propagating in general backgrounds. We have shown that these equation can be made invariant under space-time gauge transformations using the loop variable technique. The equations obtained are by construction quadratic in the fields and in this sense is similar to BRST string field theory. The main difference is that
the gauge transformation law is unchanged by the presence of interactions. The interaction terms can be written in terms of gauge invariant field strengths for the massive fields where the fact the mass is non zero is crucial.  This is reminiscent of the Born-Infeld action for the massless vector in open string theory in which all interactions involve only the field strength. We have demonstrated this explicitly  for the massive spin 2 and spin 3 fields and outlined the pattern for the next level. It is natural to conjecture that it can be done for all levels.

Since the RG method of obtaining 
equations of motion works for any background (i.e one does not need to perturb around a conformal background) this method is background independent.  Furthermore the gauge invariance does not depend on world sheet symmetries.
This means that one can easily add a UV regulator and modify the theory at intermediate stages of the calculation.
This we also know is necessary for going off shell.
This freedom also turned out to be useful for other  reasons in that we chose the extra coordinate, which played a  role similar to that of a bosonized ghost of string field theory, massive. This turned out to be necessary for the correlation functions to agree with those of string theory. 

One should add that quite independent of string theory, (at least at the tree level) this technique gives a gauge invariant massive interacting higher spin theory in any dimension. The tachyon can also be made massive if necessary. The constraint about dimension and mass spectrum comes form requiring agreement with string theory.  This presumably also ensures consistency at the loop level. If this is found to be not necessary then one can generalize to other theories.

There are many questions that need to be answered. Probably the most pressing is whether this technique can be generalized to closed strings. That would give a quadratic equation of motion, unlike string field theory, which for closed strings is non polynomial. The other pressing question is to construct an action. Another issue is to give a rigorous proof of the equivalence of the S-matrix of this theory with that of string theory.  

\appendix

\renewcommand{\thesection}{\Alph{section}}
\renewcommand{\theequation}{\thesection.\arabic{equation}}

\section{Appendix: Level 3 Quadratic Terms}
\label{appena}
\setcounter{equation}{0}

We give the calculation of the level 3 quadratic pieces:

\[
{\p L\over \p Y^\mu}= i\kom \Big[iK_3^\rho Y_3^\rho +  iK_{21}^\rho {\pp Y^\rho \over \p x_2 \p x_1} + iK_{111}^\rho {\p^3 Y^\rho\over \p x_1^3} 
-K_{11}^\rho \ki ^\sigma {\pp Y^\rho\over\p x_1^2}{\p Y^\sigma \over \p x_1}
\]
\[
 -Q_2 \kor \ki^\sigma {\p Y^\rho\over \p x_2} {\p Y^\sigma\over \p x_1} - {(i\ki.Y_1)^3\over 3!}
 \Big]\e
\]
\[
\p_{x_1}{\p L\over \p \yim}=\Big[ -K_{11}^\rho {\p^3 Y^\rho\over \p x_1^3}\kim - K_{11}^\rho {\p^2 Y^\rho\over \p x_1^2}\kim(i\ko.Y)-\]\[
Q_2\kor {\pp Y^\rho \over \p x_2 \p x_1}\kim- Q_2\ko.Y_2 \kim (i\ko.Y_1) -i\kim\ki.Y_2 \ki.Y_1 - {i\kim\over 2} (\ki.Y_1)^2(i\ko.Y_1)\Big] \e
\]
\[
\p_{x_2}{\p L\over \p \ytm}= -[Q_2 \kom\ki.Y_3 +Q_2\kom (\ki.Y_1)(i\ko.Y_2)]\e
\]
\[
\p_{x_1}^2{\p L\over \p ({\pp Y^\mu\over \p x_1^2})}=-\Big[ K_{11}^\mu \ki.Y_3 + 2 K_{11}^\mu \ki.Y_2 i\ko.Y_1 + K_{11} ^\mu \ki.Y_1 [ i\ko.Y_2 + (i\ko.Y_1)^2]\Big] \e
\]
\[
\p_{x_1}\p_{x_2}{\p L\over \p ({\pp Y^\mu\over \p x_1 \p x_2})}=[iK_{21}^\mu i\ko.Y_3 + iK_{111}^\mu i\ko.Y_2 i\ko.Y_1]\e
\]
\[
\p_{x_1}^3{\p L\over \p ({\p^3Y^\mu\over \p x_1^3})}=iK_{111}^\mu [ i\ko.Y_3 +3i\ko.Y_2 i\ko.Y_1 + (i\ko.Y_1)^3]\e
\]
\be    \label{quad3}
\p_{x_3}{\p L\over \p Y_3^\mu}=iK_3^\mu i\ko.Y_3\e
\ee
Adding these terms (with appropriate signs \footnote{Terms generated by $\p_{x_n}$ come with minus signs (as in the usual Lagrange's equations),
Terms with $\p_{x_n}\p_{x_m}$ come with a plus sign, and $\p_{x_1}^3$ comes with a minus sign}) we get
\[ L_3^\mu(z) \equiv \Big[V_3^{\mu \nu}Y_3^\nu(z) + V_{21}^{\mu \rho \sigma} Y_2^\rho(z) Y_1^\sigma(z) + V_{111}^{\mu \lambda \rho \sigma} Y_1^\lambda(z) Y_1^\rho (z)Y_1^\sigma (z)\Big] e^{i\ko.Y(z)}
\]
where
\[
V_3^{\mu \rho}=-\kom[K_3^\rho + K_{21}^\rho + K_{111}^\rho] + \kim [K_{11}^\rho + Q_2 \kor] + \kom Q_2 \kir - K_{11}^\mu \kir - K_{21}^\mu \kor + K_{111}^\mu \kor + K_3^\mu \kor
\]
\[
V_{21}^{\mu \rho\sigma} = i\Big[ -\kom K_{11}^\rho \kir +  \kim K_{11}^\rho \ko^\sigma + \kim Q_2 \kor \ko^\sigma + \kim \kir \ki^\sigma
-2 K_{11}^\mu \kir \ko^\sigma -K_{11}^\mu \kor \ki^\sigma - K_{21}^\mu\kor \ko^\sigma + 3 K_{111}^\mu \kor \ko^\sigma \Big]
\]
\[
V_{111}
^{\mu \la \rho \sigma}= 
{1\over 3!} 
\kom k_1 ^\la \kir \ki ^\sigma -
 {1\over 3!}\kim \ki ^{(\lambda} \kir \ko^{\sigma)} + 
{1\over 3} K_{11}^\mu \ki ^{(\lambda} \kor \ko^{\sigma)} - K_{111}^\mu \ko^{\lambda} \kor \ko^{\sigma}
\]

\section{Appendix: Operator Product Expansion}
\label{appenb}
\setcounter{equation}{0}

We work out some of the operator product expansions that are needed. The general master formula can be easily written in terms of
 loop variables:
\[
: e^{(i \ko .Y + i \ki .Y_1 +...+ i k_n. Y_n)(z_1)}:: e^{(i p_0 .Y + i p_1 .Y_1 +...+ i p_m. Y_m)(z_2)}:=\]
\[
e^{-\sum_{n,m} k_n.p_m G_{n,m}(z_1-z_2)} \]
\be	\label{OPE}
:e^{(i \ko .Y + i \ki .Y_1 +...+i k_n. Y_n)(z_1) +(i p_0 .Y + i p_1 .Y_1 +...+ i p_m. Y_m)(z_2)}:
\ee 
where $G_{n,m}(z_1-z_2)=\lan Y_n (z_1) Y_m(z_2)\ran$.

We can extract from (\ref{OPE}) terms multilinear in $k_i$ and $p_j$ to extract OPE's of the usual vertex operators.

The following gives the general normal ordering of vertex operators at one point:
\be   \label{NO}
e^{(i \ko .Y + i \ki .Y_1 +...+i k_n. Y_n)(z_1)} = e^{-\hf \sum_{n,m} k_n.k_m G_{n,m}(z_1,z_1)}:e^{(i \ko .Y + i \ki .Y_1 +...+ i k_n. Y_n)(z_1)}:
\ee

Thus typically one uses (\ref{NO}) followed by (\ref{OPE}) if the vertex operators are not normal ordered to begin with.

\subsection{\bf OPE of level 1 vector vertex operators:}

 The bilinear in $k_1 p_1$ gives us the OPE between $i:k_1 .Y_1 e^{i\ko. Y(z_1)}:$ 
and $i:p_1. Y_1 e^{ip_0. Y(z_2)}:$. (Thus in the situation of interest to us $\kim \yim $ would be replaced by $ F^{\nu \mu}(\ko) \yim$).

We get the following four terms: (We have suppressed the argument $(z_1-z_2)$ of the Greens function in the equations below)
\[
i)~~~~~:i k_1 .Y_1(z_1)  i p_1 .Y_1(z_2) e^{i\ko. Y(z_1)+ip_0. Y(z_2)}:
\]
\[
ii)~~~~~-\ko.p_1 G_{0,1}: i k_1 .Y_1(z_1)  e^{i\ko. Y(z_1)+ip_0. Y(z_2)}:
\]
\[
iii)~~~~~-\ki. p_0 G_{1,0}: i p_1. Y_1(z_2)  e^{i\ko. Y(z_1)+ip_0. Y(z_2)}:
\]
\[
iv)~~~~~~ \ko.p_1 p_0.\ki [G_{0,1}G_{1,0}]: e^{i\ko. Y(z_1)+ip_0. Y(z_2)}:
\]
\[
v)~~~~~~-\ki . p_1 G_{1,1} :e^{i\ko. Y(z_1)+ip_0. Y(z_2)}:
\]

Now a Taylor expansion about $z_1$ can be performed to extract various contributions. The contribution to the level 1
vertex operator $\yim (z_1)e^{i\ko. Y(z_1)}$ is:
\[
[-i \ko .p_1 \kim G_{0,1} - i p_0. \ki p_1^\mu G_{1,0} + i (z_2-z_1)G_{0,1}G_{1,0} \ko. p_1 \ki. p_0 p_0^\mu 
\]
\be 
 - {(z_2-z_1)}
\ki .p_1 G_{1,1} i p_0^\mu  ]
:\yim e^{i(\ko+p_0). Y(z_1)}:
\ee

Similarly one can extract the contribution to the level 2 vertex operator $\yim \yin e^{i\ko. Y (z_1)}$. The result is:
\[
[ \kim p_1^\nu + (z_2-z_1)\ko. p_1 G_{0,1} \kim p_0^\nu + (z_2-z_1)p_0.\ki p_1^\mu \kon G_{1,0}  +\]
\[-{(z_2-z_1)^2\over 2}G_{0,1}G_{1,0} \ki.p_0 p_1.\ko p_0^\mu p_0^\nu +{(z_2-z_1)^2\over 2} G_{1,1} \ki .p_1 p_0^\mu p_0^\nu] :\yim \yin e^{i(\ko + p_0). Y(z_1)}:
\]

\subsection{OPE of level 1 and level 2}

We have to pick terms proportional to $\kim \kin p_1^\rho$ to get the OPE of $-\hf \ki .Y_1 \ki.Y_1e^{i\ko.Y}$ and $i p_1.Y_1e^{ip_0.Y}$:

The contribution to level 1 is
\[[{(i \ki.p_0 G_{1,0})^2 \over 2} p_1^\mu + i\ki .p_0 G_{1,0} p_1.\ki G_{0,1} \kim +i(z_2-z_1){(\ki .p_0 G_{1,0})^2\over 2} p_1.\ko G_{0,1} p_0^\mu \]\[
-i \ki.p_1 G_{1,1} \kim + i (z_2-z_1)\ki .p_1 G_{1,1} \ki .p_0 G_{1,0} p_0^\mu ] :\yim e^{i(\ko + p_0). Y(z_1)}:\]

The contribution to level 2 $\yim \yin$ is
\[ 
[\ki.p_0 G_{1,0} \kim p_1^\nu - (z_2-z_1){(\ki .p_0 G_{1,0})^2\over 2} p_1^mu p_0^nu + \hf p_1.\ko G_{0,1} \kim \kin  - \]\[(z_2-z_1)\ki .p_0 p_1.\ko G_{0,1}G_{1,0} \kim p_0^\nu - {(z_2-z_1)^2\over 2} \ki.p_1 G_{1,1} \ki .p_0 G_{1,0} p_0^\mu p_0^\nu \]
\[
(z_2-z_1)^2 {(\ki .p_0 G_{1,0})^2\over 4} p_1.\ko G_{0,1} p_0^\mu p_0^\nu  + (z_2-z_1)\ki .p_1 G_{1,1} \kim p_0^\nu ] :\yim \yin e^{i(p_0+k_0).Y(z_1)}:
\]

\subsection{OPE of level 2 and level 2}

We obtain the OPE of $ :\hf (\ki. Y_1)^2 e^{i\ko.Y(z_1)}:$ and $:\hf (p_1.Y_1)^2 e^{ip_0.Y(z_2)}:$. 

We have to pick terms proportional to $\kim \kin p_1^\rho p_1 ^\sigma$.

We give the contribution to level 1:
\[
[-i {(\ki.p_0G_{1,0})^2\over 2}p_1.\ko G_{0,1} p_1^\mu -i{(p_1.k_0G_{0,1})^2\over 2}p_0.\ki G_{0,1} k_1^\mu+\]\[
{(\ki.p_0G_{1,0})^2\over 2}{(p_1.k_0G_{0,1})^2\over 2}p_0^\mu (z_2-z_1) +(z_2-z_1) {(\ki .p_1 G_{1,1})^2\over 2} i p_0^\mu + i \ki.p_1 G_{1,1} \ki .p_0 G_{1,0} i p_1^\mu \]
\[
 +i \ki.p_1 G_{1,1} \ko .p_1 G_{0,1} i k_1^\mu - \ki.p_1 G_{1,1} p_1.\ko G_{0,1} \ki.p_0 G_{1,0} (z_2-z_1) p_0^\mu 
] :\yim
e^{i(\ko +p_0).Y(z_1)}:
\]

Similarly the contribution to $\yim \yin$ consists of nine terms. We do not list them here.

\end{document}